\documentclass{SCIS2025}

\graphicspath{{figure/}}

\begin{document}

\ArticleType{RESEARCH PAPER}
\Year{2025}
\Month{January}
\Vol{68}
\No{1}
\DOI{}
\ArtNo{}
\ReceiveDate{}
\ReviseDate{}
\AcceptDate{}
\OnlineDate{}
\AuthorMark{}
\AuthorCitation{}

\title{Bayesian Probability Fusion for Multi-AP Collaborative Sensing in Mobile Networks}{Liu S H, Li X K, Huang Y M, et al. Bayesian probability fusion for multi-AP collaborative sensing in mobile networks}

\author[1,2]{Shengheng LIU}{}
\author[1]{Xingkang LI}{}
\author[1,2]{Yongming HUANG}{{huangym@seu.edu.cn}}
\author[2]{Yuan FANG}{}
\author[1,2]{Qingji JIANG}{}
\author[2]{\\Dazhuan XU}{}
\author[2]{Ziguo ZHONG}{}
\author[1,2]{Dongming WANG}{}
\author[1,2]{Xiaohu YOU}{{xhyu@seu.edu.cn}}

\address[1]{School of Information Science and Engineering, Southeast University, Nanjing {\rm 210096}, China}
\address[2]{Purple Mountain Laboratories, Nanjing {\rm 211100}, China}

\abstract{Integrated sensing and communication (ISAC) is widely acknowledged as a foundational technology for next-generation mobile networks. Compared with monostatic sensing, multi-access point (AP) collaborative sensing endows mobile networks with broader, more accurate, and resilient sensing capabilities, which are critical for diverse location-based sectors. This paper focuses on collaborative sensing in multi-AP networks and proposes a Bayesian probability fusion framework for target parameter estimation using orthogonal frequency-division multiplexing (OFDM) waveform. The framework models multi-AP received signals as probability distributions to capture stochastic observations from channel noise and scattering coefficients. Prior information is then incorporated into the joint probability density function to cast the problem as a constrained maximum \emph{a posteriori} estimation. To address the high-dimensional optimization, we develop a prior-constrained gradient ascent (PCGA) algorithm that decouples correlated parameters and performs efficient gradient updates guided by the target prior. Theoretical analysis covers optimal fusion weights for global signal-to-noise ratio maximization, PCGA convergence, and the Cram\'{e}r-Rao lower bound of the estimator, with insights applicable to broader fusion schemes. Extensive numerical simulations and real-world experiments with commercial devices show the framework reduces transmission overhead by $90\%$ versus signal fusion and lowers estimation error by $41\%$ relative to parameter fusion. Notably, field tests achieve submeter accuracy with $50\%$ probability in typical coverage of mmWave APs. These improvements highlight a favorable balance between communication efficiency and estimation accuracy for practical multi-AP sensing deployment. 

The dataset is released for research purposes and is publicly available at: http://pmldatanet.com.cn/dataapp/multimodal}


\maketitle

\section{Introduction}

Next-generation mobile networks \cite{2022ITUR, 2023YouTowards}, as the cornerstone of digital infrastructure and industrial upgrading, are undergoing rapid evolution to empower strategic emerging sectors such as low-altitude economy, automated manufacturing, and intelligent transportation \cite{2022IMT, 3GPP}. These forward-looking applications impose stringent requirements not only on quality of wireless connectivity but also on endogenous capability of physical environment perception \cite{2021YouTowards}. Against this backdrop, integrated sensing and communication (ISAC) has been advocated as a core functionality of 6G, recognized globally by initiatives such as the NextG Alliance's 6G Roadmap and the European 6G-IA's vision \cite{2024Gonz}. By sharing hardware platforms, reusing spectrum resources, and unifying signal processing chains, ISAC breaks the long-standing separation between communication and sensing systems, thereby enhancing spectrum efficiency while reducing hardware cost and deployment complexity \cite{2022LiuIntegrated}. Such integration also enables a virtuous cycle where communication links provide reliable channels for real-time transmission of sensing data, whereas sensing-derived environmental knowledge, such as channel blockage prediction and multipath characterization, reciprocally enhance communication adaptability through predictive beamforming and dynamic resource allocation \cite{TOIT25}. However, as application scenarios expand and performance requirements escalate, conventional monostatic ISAC faces insurmountable limitations. These include constrained spatial coverage, weak interference coordination, and vulnerability to blockages. To address these bottlenecks, ISAC systems with multiple cooperative access points (APs) have been proposed \cite{2025LiuFeasibility, 2024LiuISAC, 2024XuAccess}. By leveraging cooperative signal transmission, reception, and joint information processing among geographically distributed APs, networked ISAC systems significantly expand spatial sensing coverage and enables multi-perspective observation, as well as continuous target tracking in occlusion-rich scenes. Recent advances along this line include bistatic and multistatic sensing for non-line-of-sight (NLOS) detection via reflected paths, cell-free massive multipe-input-multiple-output for eliminating boundary-induced sensing blind spots, and reconfigurable intelligent surface assisted configurations that create controlled reflection channels to enhance NLOS sensing reliability \cite{2023WangFull, 2023CaoExperimental}.

A core challenge in fully unleashing the potential of multi-AP collaborative ISAC lies in \textit{fusion of sensed information}. This process involves sharing and combining data collected from distributed APs to form an accurate representation of the surveilled scene \cite{2024YangCoordinated}. Effective fusion can mitigate conflicts among multi-source data, leverage complementary information, and support mutual confirmation. Nevertheless, it is frequently hampered by uncertainties such as measurement noise inherent to individual APs and spatio-temporal asynchrony across the network. Without robust fusion mechanisms to counteract these issues, cumulative errors and observational inconsistencies inevitably arise, ultimately degrading sensing reliability.

Information fusion can be implemented at various stages of the processing chain. If sensed information is fused at a signal level, spatially separated APs transmit their raw receive signals to a global fusion center. There, waveform superposition techniques \cite{2025ZhangSignal} can be applied to integrate multi-AP signals, which directly upgrades energy gain and extends target detection range. Alternatively, each AP evaluates its local matched filtering results, i.e., the range-Doppler map, and then combines all local maps in the fusion center. This framework is known as backprojection (BP) \cite{2022AdhamNear, 2025Favarelli}, which offers a coherent processing gain and distinguishes clutter from moving objects with computational efficient fast Fourier transforms. However, both raw and processed signal fusions incur substantial overhead and demand large bandwidth for transmission. In addition, it demands nanosecond-level precise phase synchronization across all APs, which far exceeds the requirement for wireless communications at a sub-microsecond-level. This necessitates additional high-precision phase calibration schemes \cite{2025LiuModel, 2014Yin} that results in prohibitive hardware complexity and deployment costs. Therefore, many practical systems resort to fusion of tracker outputs rather than full measurement data. Nevertheless, track fusion discards useful information present in the original echoes and induces irreversible information loss during local feature extraction \cite{2023XiongDistributed, 2024VermaTrack}. To balance information preservation with implementation feasibility, many recent efforts has been focused on fusion at the intermediate levels. Parameter fusion \cite{2024MoussaMulti, 2022ZhangEfficient}, for instance, aggregates locally estimated sensing parameters such as roundtrip delay, angle of arrival (AoA), and Doppler shift from each AP instead of signals. Symbol fusion \cite{2024WeiSymbol} relaxes synchronization requirements by exploiting phase variations in demodulated symbol vectors, but is susceptible to noise amplification from multiplicative operations. Weighting strategies, such as those based on signal-to-noise ratio (SNR) \cite{2025LiuHRT} or sensor data consistency \cite{2021RenImproved}, have also been explored to optimize global detection performance.

Recent advances in probabilistic fusion have provide a unified framework for integrating multi-sensor data while quantifying inherent uncertainties. By representing sensing outputs as random finite sets \cite{2021Yi} or probability distributions \cite{2022KolianderFusion}, these methods mitigate information loss from hard decisions and accommodate heterogeneous observation qualities across nodes. For instance, grid-based likelihood ratio fusion under the maximum likelihood criterion has been applied to suppress radar ghosts and improve localization accuracy \cite{2021GaoReliable}. Similarly, Bayesian frameworks such as optimal Bayesian fusion recursively combine classification probabilities from distributed radars by leveraging conditional independence assumptions, while recursive Bayesian classification incorporates temporal updates for sequential decision-making \cite{2025Potter}. Other probabilistic forms, including maximum \emph{a posteriori} (MAP) and non-linear least squares \cite{2024FigueroaCooperative}, have been reconstructed for multi-static scenarios. However, these methods predominantly operate at the parameter level, relying on joint probability functions derived from parametric estimation processes. Consequently, they often overlook the intrinsic structural attributes of physical-layer signals, such as amplitude fluctuations, phase variations, and spatial-temporal correlations embedded in raw waveforms.

To address these gaps, we propose a Bayesian probability fusion framework specifically designed for orthogonal frequency division multiplexing (OFDM)-based mobile networks. The core innovation lies in harnessing the physical-layer structural characteristics of signals through a rigorous probabilistic characterization of raw measurements. Under a Swerling-I target fluctuation model, we derive the joint probability density function (pdf) of received signals, explicitly accounting for uncertainties arising from channel noise and scattering coefficients. This approach effectively resolves discrepancies in channel states caused by stochastic observations. Furthermore, by embedding prior target information, such as kinematic constraints and spatial priors, we formulate parameter estimation as a constrained MAP problem, which unifies observed data and prior knowledge within a Bayesian paradigm. To solve this problem efficiently, a prior-constrained gradient ascent (PCGA) algorithm is developed to decouple correlated parameters and enable scalable optimization, significantly reducing computational complexity compared to exhaustive traversal methods. The framework not only enhances fusion accuracy but also provides a statistical foundation for collaborative sensing in multi-AP networks, bridging the gap between signal-level fidelity and practical deployment constraints. The technical contributions of this work are fourfold.
\begin{itemize}
	\item We propose a Bayesian probability fusion framework that incorporates probabilistic characterization of physical-layer signals and prior target information. By leveraging structural attributes of raw signals, the framework provides a unified representation that mitigates observation heterogeneity. The constrained MAP formulation enables holistic fusion of sensed data within a consistent Bayesian paradigm and significantly reduces communication overhead.
	
	\item We develop a prior-constrained gradient-based algorithm, referred to as PCGA, to efficiently solve the high-dimensional MAP estimation. By exploiting target priors and local convexity of the fused spatial pseudo-spectrum, PCGA achieves scalable optimization with markedly lower computational complexity than exhaustive search methods.
	
	\item We establish a comprehensive theoretical foundation of the framework, including the derivation of an optimal weighting scheme from a global SNR perspective, a convergence guarantee for PCGA based on pseudo-spectrum convexity, and the Cram\'{e}r-Rao lower bound (CRLB) for theoretical estimation precision. These analytical results extend beyond the present framework to general multi-sensor fusion systems.
	
	\item We carry out extensive validation through numerical simulations and real-world experiments using commercial mmWave equipment in outdoor environments. Results confirm the framework's enhanced estimation performance (submeter accuracy w\textbackslash $50\%$ probability @$200\;{\rm{m}}$ range and $0.01\;{\rm{m^2}}$ RCS) and robustness, while complexity and overhead analyses demonstrate its practical feasibility.
\end{itemize}

\textit{Notations:} Lower (upper)-case bold characters are used to denote vectors (matrices), and the vectors are by default in column orientation. The superscripts $(\cdot)^{-1}$, $(\cdot)^{\mathsf T}$, and $(\cdot)^{\mathsf H}$ represent the inverse, transpose, and conjugate transpose operators, respectively. Symbols $\odot$ and $\otimes$ denote the Hadamard and Kronecker product operations, respectively. $\exp(\cdot)$ denotes the natural exponential function. $\left\Arrowvert\cdot\right\Arrowvert_2$ denotes the $2$-norm of vector. $\mathrm{diag}\{\cdot\}$ denotes the operation of forming a diagonal matrix. $\left\lfloor\cdot\right\rfloor$ represents the floor function. $\mathbb{E}\{\cdot\}$ returns the expected value of a discrete random variable. $\mathbb{R}$ and $\mathbb{C}$ represent the sets of real and complex numbers, respectively. ${\mathrm j}$ represents the imaginary unit.

\section{System Model}
\label{sec:model}
\vspace{-0.3cm}

\begin{figure}[]
	\vspace{-0.3cm}
	\centering
	\includegraphics[width=0.75\linewidth]{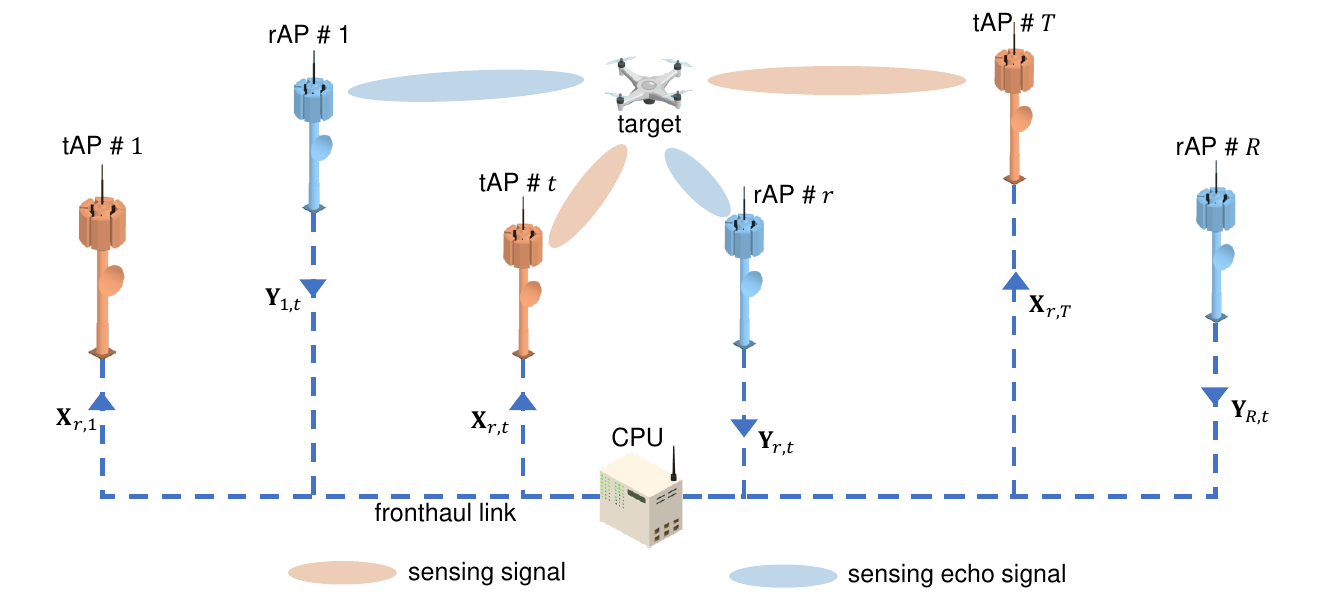}
	\vspace{-0.2cm}
	\caption{Illustration of multi-AP collaborative sensing in wireless network.}
	\label{fig:system0}
	\vspace{-0.3cm}
\end{figure}

We consider an OFDM-based multi-AP collaborative sensing network as illustrated in Fig.\ref{fig:system0}, in which $T$ transmit APs (tAPs) and $R$ receive APs (rAPs) are equipped with uniform linear array (ULA) consisting of $M$ antennas \cite{2016HeathOverview}. All APs collaboratively sensing one target. Given cost constraint, existing commercial access devices commonly use a single radio frequency (RF) chain configuration. We assume a low cost analog beamforming structure at all APs, where a single RF chain is connected to the $M$ antenna elements via phase shifters. As a result, APs lack accurate angle estimation capabilities in the scenario under consideration. Let $\mathbf{f}_{t}\in\mathbb{C}^{M\times 1}$ denote the transmit analog beamforming matrix at the $t$-th tAPs, and $\mathbf{w}_{r}\in\mathbb{C}^{M\times 1}$ denote the receive analog beamforming matrix at the $r$-th rAPs. The uniform discrete Fourier transform (DFT) beam codebook \cite{2017SuhConstruction} of size $N_{\mathrm{a}}$ for ULA with $M$ antennas is described by $\mathbf{C}=[\mathbf{c}_{1},\ldots,\mathbf{c}_{n_{\mathrm{a}}},\ldots,\mathbf{c}_{N_{\mathrm{a}}}]$, where $\mathbf{c}_{n_{\mathrm{a}}}=\frac{1}{\sqrt{M}}[1,\ldots,\mathrm{e}^{-{\mathrm j} 2\uppi (M-1) \frac{n_{\mathrm{a}}}{N_{\mathrm{a}}}}]^{\mathsf{T}}$. In particular, $\mathbf{f}_{t}$ and $\mathbf{w}_{r}$ can be selected from the predefined codebook $\mathbf{C}$ to achieve beam steering, thereby obtaining beamforming gain.

\vspace{-0.3cm}
\subsection{Receive Signal Model}
\vspace{-0.3cm}

First, we introduce the baseband signal model. Specifically, we consider the OFDM waveform used for sensing, comprising $K$ subcarriers with a subcarrier spacing of $\Delta_{\mathrm{f}}$ and operating at a carrier frequency $f_{\rm{c}}$ \cite{2023Wei5G}. A sensing signal frame consist of $L$ symbols with symbol repetition interval $T_{\mathrm{P}}$. We use Zadoff-Chu method to generate $T$ orthogonal sensing signal sequences $\{{x}_{t}\} (t\in{1,2,\ldots,T})$ for distinguishing transmitted signals from different tAPs.
As such, the received frequency-domain baseband echo signal by the $r$-th rAP transmitted from the ${t}$-th tAP at symbol $l$ on subcarrier $k$ is given by
\begin{align}\label{eq:signals} \widetilde{y}_{r,t}(k,l)=\mathbf{w}_{r}^{\mathsf{H}}\mathbf{H}_{r,t}(k,l)\mathbf{f}_{t}{x}_{t}(k,l)+\widetilde{z}(k,l),
\end{align}
where ${x}_{t}(k,l)$ represents the transmitted baseband frequency-domain OFDM symbol by the $t$-th tAP at symbol $l$ on subcarrier $k$, ${\mathbf{H}}_{r,t}(k,l)\in\mathbb{C}^{M\times M}$ represents the sensing channel, $\widetilde{z}(k,l)\sim\mathcal{CN}(0,{\sigma}^2)$ is zero mean circular symmetric complex Gaussian noise. The modulation and demodulation of the sensing signals can be deployed into wireless communication systems by reusing existing communication protocols with time division duplex mode.

\vspace{-0.3cm}
\subsection{Sensing Channel Model}
\vspace{-0.3cm}

Then, we consider the simplified Saleh-Valenzuela channel model, where the sensing channel ${\mathbf{H}}_{r,t}(k,l)$ of the tAP-target-rAP link is modeled as
\begin{equation}\label{eq:sensingchannel}
	\begin{aligned}
		{\mathbf{H}}_{r,t}(k,l)=\alpha_{r,t}\mathrm{e}^{-{\mathrm j}2\uppi k\Delta_{\mathrm{f}} \frac{d_{r,t}}{c}}\mathrm{e}^{{\mathrm j}2\uppi lT_{\mathrm{P}} \frac{v_{r,t}}{\lambda}}\times{\mathbf{a}}_{r}(\theta_{r}){\mathbf{a}}_{t}^{\mathsf{H}}(\phi_{t}),
	\end{aligned}
\end{equation}
where $c$, $\alpha_{r,t}$ and $\lambda$ represent the speed of light, reflected complex channel gains, and wavelength of target between the $r$-th rAP and the $t$-th tAP, respectively. $d_{r,t}$ and $v_{r,t}$ are the bistatic range and the corresponding speed of the target between the $r$-th rAP, and the $t$-th tAP, respectively. $\theta_{r}$ and $\phi_{t}$ represent angle of arrival at the $r$-th rAP and angle of departure at the $r$-th tAP, respectively. Assuming the antenna spacing of the ULA is half-wavelength, the steering vector for antennas of the
$r$-th rAP and the $t$-th tAP are respectively given by $\mathbf{a}_{r}(\theta_{r})=[1,\ldots,\mathrm{e}^{-{\mathrm j} \uppi (M-1) \sin(\theta_{r})}]^{\mathsf{T}}$ and $\mathbf{a}_{t}(\phi_{t})=[1,\ldots,\mathrm{e}^{-{\mathrm j} \uppi (M-1) \sin(\phi_{t})}]^{\mathsf{T}}$.

To facilitate analysis, we define a equivalent channel coefficient $\widetilde{h}_{r,t}(k,l)$ as
\begin{align}\label{eq:singlesensingchannel}
	\widetilde{h}_{r,t}(k,l)=\mathbf{w}_{r}^{\mathsf{H}}\mathbf{H}_{r,t}(k,l)\mathbf{f}_{t}=\beta_{r,t}\mathrm{e}^{-{\mathrm j}2\uppi k\Delta_{\mathrm{f}} \frac{d_{r,t}}{c}} \mathrm{e}^{{\mathrm j}2\uppi lT_{\mathrm{P}} \frac{v_{r,t}}{\lambda}},
\end{align}
where $\beta_{r,t}=\alpha_{r,t}{\widetilde{a}}_{r}{\widetilde{a}}_{t}$ is the equivalent reflection complex channel gain, and ${\widetilde{a}}_{r}=\mathbf{w}_{r}^{\mathsf{H}}\mathbf{a}_{r}(\theta_{r})$, ${\widetilde{a}}_{t}=\mathbf{f}_{t}^{\mathsf{H}}\mathbf{a}_{t}(\phi_{t})$. Note that ${\widetilde{a}}_{r}$ and ${\widetilde{a}}_{t}$ are the beamforming gains of the rAP and tAP.\footnote{The focus of this work is on parameter estimation, and beam alignment \cite{2022Chen2022} is beyond of the scope of this paper. We thus assume that since the target is within the beam main lobe, beamforming gain $G=\left|{\widetilde{a}}_{r}{\widetilde{a}}_{t}\right|$ satisfies $M/2 \leq G \leq M$.}

Let $\widetilde{\mathbf{Y}}_{r,t}\in\mathbb{C}^{K\times L}$ denote the received sensing signal matrix in which $(k,l)$-th element is $\widetilde{y}_{r,t}(k,l)$. Thus, we have
\begin{align}\label{eq:sensingsignals1}
	\widetilde{\mathbf{Y}}_{r,t}=\mathbf{\widetilde{H}}_{r,t}\odot \mathbf{X}_{t}+\widetilde{\mathbf{Z}}=\beta_{r,t}\boldsymbol{\Psi}^{\mathrm{(dv)}}_{r,t}\odot \mathbf{X}_{t}+\widetilde{\mathbf{Z}},
\end{align}
where  $\mathbf{\widetilde{H}}_{r,t}=[\widetilde{\mathbf{h}}
_{r,t}(1),\ldots,\widetilde{\mathbf{h}}
_{r,t}(l),\ldots,\widetilde{\mathbf{h}}
_{r,t}(L)]$ is the equivalent channel coefficients matrix with $\widetilde{\mathbf{h}}
_{r,t}(l)=[\widetilde{h}_{r,t}(1,l),\ldots,\widetilde{h}_{r,t}(K,l)]^{\mathsf{T}}$.
$\boldsymbol{\Psi}^{\mathrm{(dv)}}_{r,t}=\boldsymbol{\psi}^{\mathrm{(f)}}_{r,t}(\boldsymbol{\psi}^{\mathrm{(s)}}_{r,t})^{\mathsf{T}}$  is the steering matrix, with $\boldsymbol{\psi}^{\mathrm{(f)}}_{r,t}=[1,\ldots,\mathrm{e}^{-{\mathrm j}2\uppi (K-1)\Delta_{\mathrm{f}} \frac{d_{r,t}}{c}}]^{\mathsf{T}}$ and $\boldsymbol{\psi}^{\mathrm{(s)}}_{r,t}=[1,\ldots,\mathrm{e}^{{\mathrm j}2\uppi (L-1)T_{\mathrm{P}} \frac{v_{r,t}}{\lambda}}]^{\mathsf{T}}$.
$\mathbf{X}_{t}=[{\mathbf{x}_{t}}(1),\ldots,{\mathbf{x}_{t}}(l),\ldots,{\mathbf{x}_{t}}(L)]$ is a predetermined sensing signal sequences with ${\mathbf{x}_{t}}(l)=[{x}_{t}(1,l),\ldots,{x}_{t}(K,l)]^{\mathsf{T}}$.
$\widetilde{\mathbf{Z}}=[\widetilde{\mathbf{z}}(1),\ldots,\widetilde{\mathbf{z}}(l),\ldots,\widetilde{\mathbf{z}}(L)]$ is the noise matrix with $\widetilde{\mathbf{z}}(l)=[\widetilde{z}(1,l),\ldots,\widetilde{z}(K,l)]^{\mathsf{T}}$.

\section{Bayesian Probability Fusion Framework}
\label{sec:construction}

In this section, we will first establish a Bayesian probabilistic characterization of echo signals through deriving the joint posterior probability density function (pdf). Then, we transform the range and speed parameters to the position and velocity parameters. In this way, a constrained MAP problem for position and velocity estimation is formulated by leveraging observed information and prior information.

\subsection{Joint Probability Density Function Derivation}

According to least squares algorithm, we firstly unload the transmitted sensing signal sequences from $\widetilde{\mathbf{Y}}_{r,t}$ in (\ref{eq:sensingsignals1}), which can be expressed as ${y}_{r,t}(k,l)=\widetilde{y}_{r,t}(k,l)/{x}_{t}(k,l)$ \cite{2023Wei5G}. ${z}(k,l)=\widetilde{z}(k,l)/{x}_{t}(k,l)$ is also the Gaussian noise with variance ${\sigma}^2$. Thus, we have
\begin{equation}\label{eq:hypo1}
	\begin{aligned}
\mathbf{Y}_{r,t}=\beta_{r,t}\boldsymbol{\Psi}^{\mathrm{(dv)}}_{r,t}+\mathbf{Z},
	\end{aligned}
\end{equation}
where  $\mathbf{{Y}}_{r,t}=[{\mathbf{y}}
_{r,t}(1),\ldots,{\mathbf{y}}
_{r,t}(l),\ldots,{\mathbf{y}}
_{r,t}(L)]$ is the received signal with ${\mathbf{y}}
_{r,t}(l)=[{y}_{r,t}(1,l),\ldots,{y}_{r,t}(K,l)]^{\mathsf{T}}$. ${\mathbf{Z}}=[{\mathbf{z}}(1),\ldots,{\mathbf{z}}(l),\ldots,{\mathbf{z}}(L)]$ is the noise matrix with ${\mathbf{z}}(l)=[{z}(1,l),\ldots,{z}(K,l)]^{\mathsf{T}}$. By vectorizing $\mathbf{Y}_{r,t}$, $\mathbf{Z}$ and $\boldsymbol{\Psi}^{\mathrm{(dv)}}_{r,t}$ into $\mathbf{y}_{r,t}\in\mathbb{C}^{KL\times 1}$, $\mathbf{z}\in\mathbb{C}^{KL\times 1}$ and $\boldsymbol{\psi}^{\mathrm{(dv)}}_{r,t}\in\mathbb{C}^{KL\times 1}$, the received echo signal is rewritten as
\begin{equation}\label{eq:hypo2}
	\begin{aligned} \quad\mathbf{y}_{r,t}=\beta_{r,t}\boldsymbol{\psi}^{\mathrm{(dv)}}_{r,t}+\mathbf{z}.
	\end{aligned}
\end{equation}
The sensing task focuses on the estimation of $\mathbf{d}$ and $\mathbf{v}$ from $\{\mathbf{{y}}_{1,1},\ldots,\mathbf{{y}}_{R,T}\}$, where $\mathbf{d}=\{d_{1,1},\ldots,d_{R,T}\}$ and $\mathbf{v}=\{v_{1,1},\ldots,v_{R,T}\}$ are range and speed of all t/rAP pairs, respectively.

Under the Swerling-I target fluctuation model \cite{1960SwerlingProbability}, the power of radar cross section (RCS) is given by an exponential distribution, and the RCS $\gamma$ $\sim\mathcal{CN}(0,\bar{\sigma}^2)$.
Under this assumption, we have equivalent reflected complex channel gain $\beta_{r,t}\sim\mathcal{CN}(0,\tilde{\sigma}_{r,t}^2)$, namely \cite{2024ZhaoBayesian}
\begin{equation} \beta_{r,t}=\alpha_{r,t}{\widetilde{a}}_{r}{\widetilde{a}}_{t}=\sqrt{\frac{P_\mathrm{T} G^2 \lambda^2}{(4 \uppi)^3 d^2_r d^2_t}}\gamma ,\quad \tilde{\sigma}_{r,t}^2={\frac{P_\mathrm{T} G^2 \bar{\sigma}^2 \lambda^2}{(4 \uppi)^3 d^2_r d^2_t}},
\end{equation}
where $P_\mathrm{T}$ and $G=\left|{\widetilde{a}}_{r}{\widetilde{a}}_{t}\right|$ are the transmit power and the beamforming gain, respectively. $d_r$ and $d_t$ denote the range from the $r$-th rAP to target and from target to the $t$-th tAP, respectively.

By defining $\boldsymbol{\theta}_{r,t}=\{\beta_{r,t},d_{r,t},v_{r,t}\}$, the conditional pdf of $\mathbf{y}_{r,t}$ with respect to $\boldsymbol{\theta}_{r,t}$ is given by
\begin{equation}\label{eq:pdf1}
	\begin{aligned}
		p(\mathbf{y}_{r,t}|\boldsymbol{\theta}_{r,t})=\frac{1}{\uppi^{KL} |\det(\mathbf{R}_{{r,t}})|}\exp\left(\mathbf{y}_{r,t}^{\mathsf{H}}\mathbf{R}_{{r,t}}^{-1}\mathbf{y}_{r,t}\right),
	\end{aligned}
\end{equation}
where $\mathbf{R}_{{r,t}}$ is the covariance matrix of the received signals. For particular parameter units $d_{r,t}$ and $v_{r,t}$, we calculate the covariance matrix $\mathbf{R}_{{r,t}}$ is derived as
\begin{align}\label{eq:R} \mathbf{R}_{{r,t}}=\mathbb{E}_{\beta_{r,t},\mathbf{z}}\{\mathbf{y}_{r,t}\mathbf{y}_{r,t}^{\mathsf{H}}\}=\tilde{\sigma}_{r,t}^2\boldsymbol{\psi}^{\mathrm{(dv)}}_{r,t}(\boldsymbol{\psi}^{\mathrm{(dv)}}_{r,t})^{\mathsf{H}}+\sigma^2\mathbf{I}_{KL\times KL}.
\end{align}
Then, by using Sherman-Morrison formula \cite{1950Sherman}, we have $\det(\mathbf{R}_{{r,t}})=\sigma^2+KL\tilde{\sigma}_{r,t}^2$, and $\mathbf{R}_{{r,t}}^{-1}=\frac{1}{\sigma^2}(\mathbf{I}_{KL\times KL}$ $-\frac{\tilde{\sigma}_{r,t}^2\boldsymbol{\psi}^{\mathrm{(dv)}}_{r,t}(\boldsymbol{\psi}^{\mathrm{(dv)}}_{r,t})^{\mathsf{H}}}{\sigma^2+KL\tilde{\sigma}_{r,t}^2} )$.
Thus, the conditional pdf in (\ref{eq:pdf1}) is rewritten as
\begin{align}\label{eq:pdfH1}
	p(\mathbf{y}_{r,t}|d_{r,t},v_{r,t})=\frac{1}{\uppi^{KL} (\sigma^2+KL\tilde{\sigma}_{r,t}^2)}\exp\left(-\frac{||\mathbf{y}_{r,t}||^{2}_2}{\sigma^2}+\frac{\tilde{\sigma}_{r,t}^2|(\boldsymbol{\psi}^{\mathrm{(dv)}}_{r,t})^{\mathsf{H}}\mathbf{y}_{r,t}|^2}{\sigma^2(KL\tilde{\sigma}_{r,t}^2+\sigma^2)}\right).
\end{align}
Assuming that the noise and scattering coefficients among various APs is independently, the joint pdf $p(\mathbf{{y}}_{1,1},\ldots,\mathbf{{y}}_{R,T}|\mathbf{d},\mathbf{v})$ is expressed as
\begin{align}\label{eq:pdfall}
	p(\mathbf{{y}}_{1,1},\ldots,\mathbf{{y}}_{R,T}|\mathbf{d},\mathbf{v})=\prod_{r=1, t=1}^{r=R, t=T}p(\mathbf{y}_{r,t}|{d}_{r,t},{v}_{r,t}).
\end{align}
According to Bayes' theorem, the joint posterior pdf $p(\mathbf{d},\mathbf{v}|\mathbf{{y}}_{1,1},\ldots,\mathbf{{y}}_{R,T})$ is given by
\begin{align}\label{eq:Mpdf}
	p(\mathbf{d},\mathbf{v}|\mathbf{{y}}_{1,1},\ldots,\mathbf{{y}}_{R,T})=\frac{p(\mathbf{{y}}_{1,1},\ldots,\mathbf{{y}}_{R,T}|\mathbf{d},\mathbf{v})p(\mathbf{d})p(\mathbf{v})}{	p(\mathbf{{y}}_{1,1},\ldots,\mathbf{{y}}_{R,T})},
\end{align}
where $p(\mathbf{d})$ and $p(\mathbf{v})$ are the prior probability distribution with respect to $\mathbf{d}$ and $\mathbf{v}$, respectively. As such, we define the log-MAP estimation problem as
\begin{align}\label{eq:mape1}
	\text{(P1)}:\ \arg\max_{{\mathbf{d}},{\mathbf{v}}}\ln(p(\mathbf{d},\mathbf{v}|\mathbf{{y}}_{1,1},\ldots,\mathbf{{y}}_{R,T})).
\end{align}
By substituting equations (\ref{eq:pdfH1}) and (\ref{eq:pdfall}) into (\ref{eq:Mpdf}), problem (P1) is transformed into
\begin{align}\label{eq:mape2}
	\text{(P2)}:\ \arg\max_{{\mathbf{d}},{\mathbf{v}}}\sum_{r=1,t=1}^{r=R,t=T}\frac{\rho_{r,t}^2|(\boldsymbol{\psi}^{\mathrm{(dv)}}_{r,t})^{\mathsf{H}}\mathbf{y}_{r,t}|^2}{\sigma^2(KL\rho_{r,t}^2+1)}+\ln\left(p(\mathbf{d})p(\mathbf{v})\right)-B,
\end{align}
where $\rho_{r,t}^2={\tilde{\sigma}_{r,t}}/{\sigma^2}$
is defined as the received SNR of the $t,r$-th t/rAP pair, and $B=\sum_{r=1,t=1}^{r=R,t=T}({||\mathbf{y}_{r,t}||^{2}_2}/{\sigma^2}$ $+\ln\left(\uppi^{KL} (\sigma^2+KL\tilde{\sigma}_{r,t}^2)\right))$. $\boldsymbol{\psi}^{\mathrm{(dv)}}_{r,t}$ is the steering vector of $\mathbf{d}$ and $\mathbf{v}$.

\vspace{-0.3cm}
\subsection{Estimated Parameter Conversion}
\vspace{-0.3cm}

To map the range parameters into position parameters, it is necessary to perform parameter conversion. Based on the scattering geometric relationship in $2$-dimensional ($2$D) space, the ranges of the tAP-target-rAP scattered paths can be expressed as
\begin{equation}\label{eq:BistaticDistance}
	\begin{aligned} d_{0,r,t}=d_{0,r}+d_{0,t}=||\mathbf{p}_0-\mathbf{p}_r||_2+||\mathbf{p}_0-\mathbf{p}_t||_2,
	\end{aligned}
\end{equation}
where $\mathbf{p}_0=(x_0,y_0)^{\mathsf{T}}$, $\mathbf{p}_r=(x_{r},y_{r})^{\mathsf{T}}$, $\mathbf{p}_t=(x_{t},y_{t})^{\mathsf{T}}$ denote the $2$D Cartesian coordinates of the target, the $r$-th rAP and the $t$-th tAP, respectively. The corresponding speed can be rewritten by the respective time-derivatives as
\begin{equation}\label{eq:Bistaticrate}
	\begin{aligned} v_{0,r,t}=\dot{d}_{0,r}+\dot{d}_{0,t}=\frac{(\mathbf{p}_0-\mathbf{p}_r)^{\mathsf{T}}\mathbf{v}_{\mathrm{P},0}}{||\mathbf{p}_0-\mathbf{p}_r||_2}+\frac{(\mathbf{p}_0-\mathbf{p}_t)^{\mathsf{T}}\mathbf{v}_{\mathrm{P},0}}{||\mathbf{p}_0-\mathbf{p}_t||_2},
	\end{aligned}
\end{equation}
where $\dot{d}_{0,r}$ and $\dot{d}_{0,t}$ are the change rate of  $d_{0,r}$ and  $d_{0,t}$ respectively. $\mathbf{v}_{\mathrm{P},0}=(v_{x,0},v_{y,0})^{\mathsf{T}}$ denote the velocity of target in the $\mathrm{x}-$, $\mathrm{y}-$ axis directions.
According to (\ref{eq:BistaticDistance}) and (\ref{eq:Bistaticrate}), the parameter conversion can be expressed as
\begin{subequations}\label{eq:backprojection}
	\begin{align}
		&d_{r,t}\rightarrow d_{r,t}(\mathbf{p})=||\mathbf{p}-\mathbf{p}_r||_2+||\mathbf{p}-\mathbf{p}_t||_2,\\
		&v_{r,t}\rightarrow v_{r,t}(\mathbf{p},\mathbf{v}_{\mathrm{P}})=\frac{(\mathbf{p}-\mathbf{p}_r)^{\mathsf{T}}\mathbf{v}_{\mathrm{P}}}{||\mathbf{p}-\mathbf{p}_r||_2}+\frac{(\mathbf{p}-\mathbf{p}_t)^{\mathsf{T}}\mathbf{v}_{\mathrm{P}}}{||\mathbf{p}-\mathbf{p}_t||_2},
	\end{align}
\end{subequations}
where $\mathbf{p}=(x, y)^{\mathsf{T}}$ and $\mathbf{v}_{\mathrm{P}}=(v_x,v_y)^{\mathsf{T}}$ are the position and velocity, respectively.
Therefore, we can rewrite parameter estimation problem as
\begin{equation}
	\begin{aligned}\label{eq:mapfusion}
		\text{(P3)}:\ \arg\max_{{\mathbf{p}},{\mathbf{v}}_{\mathrm{P}}}\sum_{r=1,t=1}^{r=R,t=T}\frac{\rho_{r,t}^2|(\boldsymbol{\psi}^{\mathrm{(xy)}}_{r,t})^{\mathsf{H}}\mathbf{y}_{r,t}|^2}{\sigma^2(KL\rho_{r,t}^2+1)}+\ln\left(p(\mathbf{p})p(\mathbf{v}_{\mathrm{P}})\right)-B,
	\end{aligned}
\end{equation}
where
$\boldsymbol{\psi}^{\mathrm{(xy)}}_{r,t}=\boldsymbol{\psi}^{\mathrm{(xyf)}}_{r,t}\otimes\boldsymbol{\psi}^{\mathrm{(xys)}}_{r,t}$ is the steering matrix after parameter conversion, and $\boldsymbol{\psi}^{\mathrm{(xyf)}}_{r,t}$ $=[1,\cdots,$ $\mathrm{e}^{-{\mathrm j}2\uppi K\Delta_{\mathrm{f}} \frac{d_{r,t}(\mathbf{p})}{c}}]^{\mathsf{T}}$, $\boldsymbol{\psi}^{\mathrm{(xys)}}_{r,t}=[1,\cdots, \mathrm{e}^{{\mathrm j}2\uppi L T_{\mathrm{P}} \frac{v_{r,t}(\mathbf{p},\mathbf{v}_{\mathrm{P}})}{\lambda}}]^{\mathsf{T}}$. As a result, the original estimated parameters ${\mathbf{d}}$ and ${\mathbf{v}}$ are replaced by $\mathbf{p}$ and $\mathbf{v}_{\mathrm{P}}$, respectively. Through conversion, we have reduced the number of estimation parameters from $2RT$ to $4$.

\vspace{-0.3cm}
\subsection{Estimation Problem Reformulation}
\vspace{-0.3cm}

The Bayesian probability fusion is implemented through multiplicative operation of posterior probability density functions from multiple APs. It is observed from (\ref{eq:mapfusion}) that estimated parameter-dependent term is given by
\begin{equation}
	\begin{aligned}\label{eq:fpvp} f({\mathbf{p}},{\mathbf{v}}_{\mathrm{P}})=\underbrace{\sum_{r=1,t=1}^{r=R,t=T}\frac{\rho_{r,t}^2|(\boldsymbol{\psi}^{\mathrm{(xy)}}_{r,t})^{\mathsf{H}}\mathbf{y}_{r,t}|^2}{\sigma^2(KL\rho_{r,t}^2+1)}}_{\mathrm{observed\ information}}+\underbrace{\ln\left(p(\mathbf{p})p(\mathbf{v}_{\mathrm{P}})\right)}_{\mathrm{prior\ information}},
	\end{aligned}
\end{equation}
which is the fused spacial pseudo-spectrum and comprises two components: observed information and prior information. In practice, due to the limited sensing range of the system, regions outside the sensing range are not of concern. Thus, these parameters can be assumed to follow a uniform distribution over a bounded interval, i.e.,
\begin{align}\label{eq:opt0}
		p({\mathbf{p}})=\left\{
		\begin{aligned}
			\frac{1}{\mathrm{Area}(\mathbb{P})}&,\quad \mathbf{p}\in \mathbb{P}; \\
			0&,\quad \text{otherwise};
		\end{aligned}\right.\qquad
		p({\mathbf{v}}_{\mathrm{P}})=\left\{
		\begin{aligned}
			\frac{1}{\mathrm{Area}(\mathbb{V}_{\mathrm{P}})}&,\quad {\mathbf{v}}_{\mathrm{P}}\in \mathbb{V}_{\mathrm{P}}; \\
			0&,\quad \text{otherwise},
		\end{aligned}\right.
\end{align}
where $\mathrm{Area}(\cdot)$ denotes the area of the region. $\mathbb{P}$ and $\mathbb{V}_{\mathrm{P}}$ are the position space and velocity space, respectively. Based on parameter estimation problem (P3), a constrained MAP is reformulated as
\begin{equation}
	\begin{aligned}\label{eq:opt}
		\text{(P4)}:\  \arg\max_{{\mathbf{p}},{\mathbf{v}}_{\mathrm{P}}}\sum_{r=1,t=1}^{r=R,t=T}\frac{\rho_{r,t}^2|(\boldsymbol{\psi}^{\mathrm{(xy)}}_{r,t})^{\mathsf{H}}\mathbf{y}_{r,t}|^2}{\sigma^2(KL\rho_{r,t}^2+1)}\quad\text{s.t.}\quad\mathbf{p}\in\mathbb{P},\quad {\mathbf{v}}_{\mathrm{P}}\in \mathbb{V}_{\mathrm{P}}.	
	\end{aligned}
\end{equation}
Notice that the prior information of parameter pdfs is incorporated into the constraints of estimated problem in (P4) as the constraints.\footnote{The target prior distribution can also be assumed to follow alternative forms, and additional regularization constraints can be incorporated into (P4).}
\begin{remark}
	From signal perspective, observed information can be regarded as the SNR-aware weighted square sum of the matched filtering results, and the weights are set by $\frac{\rho_{r,t}^2}{\sigma^2(KL\rho_{r,t}^2+1)}$.
	In Section \ref{sec:AnalysisA}, we prove the optimality of the weighting scheme from the perspective of global SNR.
\end{remark}
\begin{remark}
	From application perspective, under the condition that APs possess distributed processing capability, the priori information can guide the screening of raw data to reduce transmission overhead.
	\end{remark}
\begin{remark}
	Alternatively, prior information of $\mathbf{p}$ and ${\mathbf{v}}_{\mathrm{P}}$ can be assumed to follow a uniform distribution over the entire space \cite{2020ZhouSpatial}. This assumption causes the MAP estimation to degenerate into MLE \cite{2022SakhniniTarget}, and it can be applied when prior information is unavailable.
\end{remark}
Typically, such problem can be tackled using a grid-based traversal algorithm. However, the vast dimensionality of the variable space results in prohibitively high computational complexity. In the following, we will develop efficient algorithms for parameter estimation within the Bayesian probability fusion framework.

\section{Solution to Estimation Problem (P4)}
\label{sec:Solution}
\vspace{-0.3cm}

In this section, we propose a PCGA algorithm inspired by \cite{2025YangGradient}, to reduce the computational complexity of the estimation problem (P4). Specifically, A gradient-based search strategy is devised to minimize the scope of the search process. We first decouple the variables, namely $\mathbf{p}$ and  $\mathbf{v}_{\mathrm{P}}$, and optimize them separately. Then, the coordinate update is utilized to refine coordinates toward the global maximum under the constraints of target prior distribution.

\vspace{-0.3cm}
\subsection{Decoupling of Estimation Problem}
\vspace{-0.3cm}

Note that the variables in (\ref{eq:opt}) reveals that $\mathbf{p}$ and $\mathbf{v}_{\mathrm{P}}$ can be decoupled from $\boldsymbol{\psi}^{\mathrm{(xy)}}_{r,t}$. This enables us to perform decoupling by exploiting the physical significance embedded in the phase information.

A critical observation is that the position $\mathbf{p}$ is manifested in the phase component corresponding to the subcarrier index $k$, whereas the velocity $\mathbf{v}_{\mathrm{P}}$ is encoded in the phase component linked to the symbol index $l$.
This intrinsic separation enables us to decouple the estimation process: we first aggregate measurements across symbol dimensions for $\mathbf{p}$ estimation and across subcarrier dimensions for $\mathbf{v}_{\mathrm{P}}$ estimation. Based on this observation, we decompose the original estimation problem (P4) into two sub-problems as
\begin{subequations}
	\begin{align}	\text{(P4.1)}:\ &\arg\max_{{\mathbf{p}}}f_1({\mathbf{p}})\qquad\text{s.t.}\quad\mathbf{p}\in\mathbb{P},\\\text{(P4.2)}:\ &\arg\max_{{\mathbf{v}}_{\mathrm{P}}}f_2({\mathbf{v}}_{\mathrm{P}},\hat{\mathbf{p}})\qquad\text{s.t.}\quad {\mathbf{v}}_{\mathrm{P}}\in \mathbb{V}_{\mathrm{P}},
	\end{align}
\end{subequations}
where
\begin{subequations}
\begin{align}	f_1({\mathbf{p}})=\arg\max_{{\mathbf{p}}}\sum_{r=1,t=1}^{r=R,t=T}\sum_{l=1}^{l=L}\frac{\rho_{r,t}^2|(\boldsymbol{\psi}^{\mathrm{(xyf)}}_{r,t})^{\mathsf{H}}\mathbf{y}_{r,t}^{\mathrm{f}}(l)|^2}{\sigma^2(K\rho_{r,t}^2+1)}, \end{align}
\begin{align}
f_2({\mathbf{v}}_{\mathrm{P}},\hat{\mathbf{p}})=\arg\max_{{\mathbf{v}}_{\mathrm{P}}}\sum_{r=1,t=1}^{r=R,t=T}\sum_{k=1}^{k=K}\frac{\rho_{r,t}^2|(\boldsymbol{\psi}^{\mathrm{(xys)}}_{r,t})^{\mathsf{H}}\mathbf{y}_{r,t}^{\mathrm{s}}(k)|^2}{\sigma^2(L\rho_{r,t}^2+1)}.
\end{align}
\end{subequations}
Vector $\mathbf{y}_{r,t}^{\mathrm{f}}(l)=[{y}_{r,t}(1,l),\ldots,{y}_{r,t}(K,l)]^{\mathsf{T}}$ represents single-symbol signal of all subcarriers, and $\mathbf{y}_{r,t}^{\mathrm{s}}(k)=[{y}_{r,t}(k,1),\ldots,$ ${y}_{r,t}(k,L)]^{\mathsf{T}}$ is single-carrier signal of all symbols. Vector $\boldsymbol{\psi}^{\mathrm{(xyf)}}_{r,t}=[1,\cdots,\mathrm{e}^{-{\mathrm j}2\uppi K\Delta_{\mathrm{f}} \frac{d_{r,t}(\mathbf{p})}{c}}]^{\mathsf{T}}$ denotes the frequency-domain steering vector, and $\boldsymbol{\psi}^{\mathrm{(xys)}}_{r,t}=[1,\cdots, \mathrm{e}^{{\mathrm j}2\uppi L T_{\mathrm{P}} \frac{v_{r,t}(\hat{\mathbf{p}},\mathbf{v}_{\mathrm{P}})}{\lambda}}]^{\mathsf{T}}$ denotes the time-domain steering vector.

Note that the objective function in problem (P4.2) depends on $\hat{\mathbf{p}}$. Thus, the solution to problem (P4.1) is used as input in solving problem (P4.2). This decoupling strategy reduces the search dimensionality and facilitates the analysis of the spectral properties in the position space and velocity space. In the following, we will introduce the algorithm for solving subproblem problem (P4.1) for $\mathbf{p}$.

\vspace{-0.3cm}
\subsection{Prior-Constrained Grid Preset}
\vspace{-0.3cm}

A limited number of grid points can be preset to reduce the search complexity under the constraints of target prior distribution. First, we discretize this sub-problem by selecting a subset $\mathbb{P}_{\rm{g}}\subset\mathbb{P}$ composed of equally spaced points $\{\mathbf{p}_1,\ldots,\mathbf{p}_n,\ldots,\mathbf{p}_N\}$. The discretized grid points $\mathbf{p}_n$ can be represented as
\begin{align}\label{eq:mapde1}
	\mathbf{p}_n=(x_n,y_n)^{\mathsf{T}}=(x_1+n\Delta x,y_1+n\Delta y)^{\mathsf{T}},
\end{align}
where $\Delta x$, $\Delta y$ are the grid intervals in the position space. These intervals can be adjusted based on the specific region of interest. In the single-target scenario, we simply select the position $\hat{\mathbf{p}}_{\rm{c}}$ that maximize the function $f_1({\mathbf{p}})$ as the coarse estimates. This process can be formulated as
\begin{equation}\label{eq:optco}
	\begin{aligned}	\arg\max_{{\mathbf{p}}_n}f_1({\mathbf{p}}_n)\qquad\text{s.t.}\quad\mathbf{p}_n\in\mathbb{P}_{\rm{g}}.
	\end{aligned}
\end{equation}
The sparsity inherent to position space spectra imposes critical constraints on grid interval selection. Oversized intervals risk missing spectral maxima during coarse estimation.
\begin{figure*}[]
	\vspace{-0.2em}
	\centering
	\includegraphics[width=0.99\linewidth]{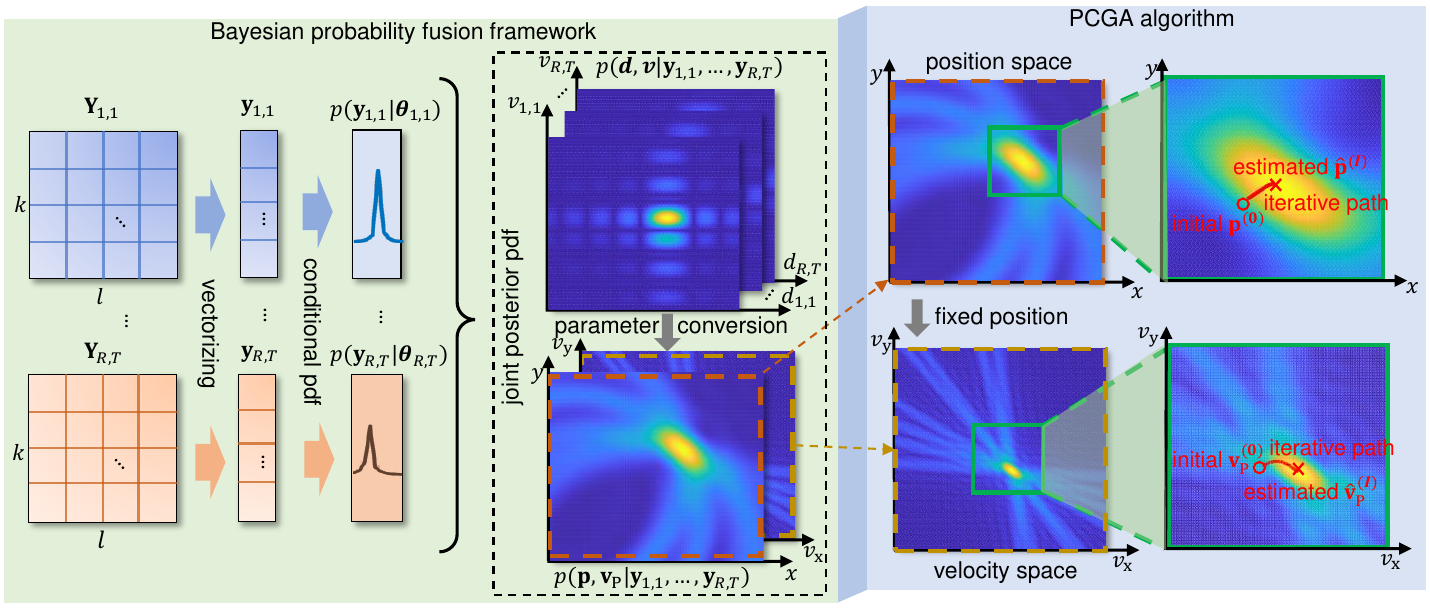}
	\vspace{-0.2cm}
	\caption{Workflow of the proposed Bayesian probability fusion framework and PCGA algorithm for single target scenario.}
	\label{fig:DRCGA}
	\vspace{-0.3cm}
\end{figure*}

\vspace{-0.3cm}
\subsection{Gradient-based Refinement}
\vspace{-0.3cm}

Then, we implement the gradient-based fine search called coordinate gradient ascent (CGA) algorithm to achieve higher precision. Specifically, we need to search for the variable that maximizes the objective function in problem (P4.1) as the estimation of the position. First and foremost, it is essential to configure suitable initial values for the iterative searching process of the CGA algorithm. Through the coarse estimation, we can determine that the target position is within the neighborhood of $\hat{\mathbf{p}}_{\rm{c}}$.

Therefore, the result $\hat{\mathbf{p}}_{\rm{c}}$ of the coarse estimation is the position in $\mathbb{P}_{\rm{g}}$ that is closest to the optimal solution and can be set as the initial value, i.e., $\mathbf{p}^{(0)}=\hat{\mathbf{p}}_{\rm{c}}$. Then, we can update the coordinates during the iteration. At the $i$-th iteration, we update $\mathbf{p}^{(i)}=(x^{(i)},y^{(i)})$ according to the following formulas
\begin{align}\label{eq:CGA1}	
		x^{(i)}=x^{(i-1)}+\eta\frac{\partial f_1(\mathbf{p}^{(i-1)})}{\partial x}\quad {\textrm{and}}\quad
		y^{(i)}=y^{(i-1)}+\eta\frac{\partial f_1(\mathbf{p}^{(i-1)})}{\partial y},
\end{align}
where $\eta$ is the learning rate that determines the step size of each update to ensure stable convergence and efficient search. Note that $f_1(\mathbf{p})$ is intricate, and it is hard to obtain its closed-form partial derivatives. To handle this issue, we adopt the finite difference method to calculate the exact numerical approximation of the partial derivatives, which are computed as
\begin{align}\label{eq:CGA}	
		\frac{\partial f_1(\mathbf{p})}{\partial x}\approx\frac{f_1({x}+\Delta,y)-f_1({x}-\Delta,y)}{2\Delta} \quad {\textrm{and}}\quad
		\frac{\partial f_1(\mathbf{p})}{\partial y}\approx\frac{f_1({x},y+\Delta)-f_1({x},y-\Delta)}{2\Delta},
\end{align}
where $\Delta$ is a extremely small value. Finally, the search terminates either upon reaching the maximum number of iterations $I$, or when the change in coordinates falls below the threshold $\epsilon$. The algorithm for solving subproblem (P4.2) follows a similar process and thus we omit the detailed introduction here. Finally, the Bayesian probability fusion framework is illustrated in Fig.~\ref{fig:DRCGA}.

\section{Performance Analysis}
\label{sec:Analysis}
\vspace{-0.3cm}

In this section, we provide several theoretical analyses in terms of the optimal weights for global SNR maximization, the convergence of the PCGA algorithm, and the CRLB of parameters estimation. These theoretical analyses demonstrate the effectiveness of the proposed fusion framework.

\vspace{-0.3cm}
\subsection{Optimal Weights for Global SNR Maximization}
\label{sec:AnalysisA}
\vspace{-0.3cm}

According to (\ref{eq:opt}), it can be known that the operation of the proposed fusion is based on the weighted square sum of the matched filtering results of all t/rAP pairs, and the weights are set by $\frac{\rho_{r,t}^2}{\sigma^2(KL\rho_{r,t}^2+1)}$.

First, we normalize the spatial pseudo-spectrum after matched filtering $|(\boldsymbol{\psi}^{\mathrm{(xy)}}_{r,t})^{\mathsf{H}}\mathbf{y}_{r,t}|^2$. By analyzing the energy distribution of the spectrum associated with the target position and velocity, we can determine the maximum value of the spectrum as
\begin{align}
	\label{eq:amplitude} E^2_{r,t}&=\max_{{\mathbf{p}},{\mathbf{v}}_{\mathrm{P}}}\mathbb{E}_{\beta_{r,t},\mathbf{z}}\left\{|(\boldsymbol{\psi}^{\mathrm{(xy)}}_{r,t})^{\mathsf{H}}\mathbf{y}_{r,t}|^2\right\}=\mathbb{E}_{\beta_{r,t},\mathbf{z}}\left\{|(\check{\boldsymbol{\psi}}^{\mathrm{(xy)}}_{r,t})^{\mathsf{H}}(\beta_{r,t} \check{\boldsymbol{\psi}}_{r,t}^{\mathrm{(xy)}} +\mathbf{z})|^2\right\}\notag=\sigma^2(KL\rho_{r,t}^2+1),	
\end{align}
where $\check{\boldsymbol{\psi}}^{\mathrm{(xy)}}_{r,t}$ corresponds to the matched filtering cell associated with the target's position and velocity. Then, let $\tilde{Y}({\mathbf{p}},{\mathbf{v}}_{\mathrm{P}})$ denotes the magnitude of the fused spectrum for particular parameter unit $\mathbf{p}$ and ${\mathbf{v}}_{\mathrm{P}}$, which can be written as $\tilde{Y}({\mathbf{p}},{\mathbf{v}}_{\mathrm{P}})=\boldsymbol{\xi}\boldsymbol{\Upsilon}({\mathbf{p}},\mathbf{v}_{\mathrm{P}})$, where $\boldsymbol{\xi}=[\xi_{1,1},\ldots\xi_{r,t},\ldots,\xi_{R,T}]$ denotes the fusion weights, $\boldsymbol{\Upsilon}({\mathbf{p}},\mathbf{v}_{\mathrm{P}})=[{\Upsilon}_{1,1}({\mathbf{p}},\mathbf{v}_{\mathrm{P}}),\cdots,{\Upsilon}_{R,T}({\mathbf{p}},\mathbf{v}_{\mathrm{P}})]^{\mathsf{T}}$ denotes the normalized spectrum and ${\Upsilon}_{r,t}({\mathbf{p}},\mathbf{v}_{\mathrm{P}})=|(\boldsymbol{\psi}^{\mathrm{(xy)}}_{r,t})^{\mathsf{H}}\mathbf{y}_{r,t}|^2/E^2_{r,t}$.

In the following, we will prove that the design of weights is optimal in terms of global SNR, which is defined as
\begin{equation}
	\label{eq:power} \rho^2=\frac{{\sum_{r=1,t=1}^{r=R,t=T}\xi_{r,t}\mathbb{E}_{\beta_{r,t},{\mathbf{p}},{\mathbf{v}}_{\mathrm{P}}}\left\{|(\boldsymbol{\psi}^{\mathrm{(xy)}}_{r,t})^{\mathsf{H}}\beta_{r,t} \check{\boldsymbol{\psi}}_{r,t}^{\mathrm{(xy)}} |^2\right\}/E^2_{r,t}}}{{\sum_{r=1,t=1}^{r=R,t=T}\xi_{r,t}\mathbb{E}_{\mathbf{z},{\mathbf{p}},{\mathbf{v}}_{\mathrm{P}}}\left\{|(\boldsymbol{\psi}^{\mathrm{(xy)}}_{r,t})^{\mathsf{H}} {\mathbf{z}} |^2\right\}/E^2_{r,t}}},
\end{equation}
where the globe signal power of single t/rAP pair is expressed as $\mathbb{E}_{\beta_{r,t},{\mathbf{p}},{\mathbf{v}}_{\mathrm{P}}}\left\{|(\boldsymbol{\psi}^{\mathrm{(xy)}}_{r,t})^{\mathsf{H}}\beta_{r,t} \check{\boldsymbol{\psi}}_{r,t}^{\mathrm{(xy)}} |^2\right\}=\tilde{\sigma}_{r,t}^2$, and the globe noise power of single t/rAP pair is expressed as $\mathbb{E}_{\mathbf{z},{\mathbf{p}},{\mathbf{v}}_{\mathrm{P}}}\left\{|(\boldsymbol{\psi}^{\mathrm{(xy)}}_{r,t})^{\mathsf{H}} {\mathbf{z}} |^2\right\}=\sigma^2$. Thus, the global SNR maximization problem after normalized spectrum fusion is formulated as
\begin{equation}
	\begin{aligned}\label{eq:power}
		\text{(P5)}:\ \arg \max_{\boldsymbol{\xi}}\frac{{\sum_{r=1,t=1}^{r=R,t=T}\xi_{r,t}\tilde{\sigma}_{r,t}^2/E^2_{r,t}}}{\sum_{r=1,t=1}^{r=R,t=T}\xi_{r,t}\sigma^2/E^2_{r,t}}.
	\end{aligned}	
\end{equation}
By leveraging the Lagrange multiplier method, the optimal solution to problem (P5) is  $\xi_{r,t} \propto\tilde{\sigma}_{r,t}^2/\sigma^2=\rho_{r,t}^2$. Under the optimal weights, fused spatial pseudo-spectrum corresponding to maximum global SNR is given by
\begin{equation}
	\begin{aligned}\label{eq:weightfusion}
		\tilde{Y}({\mathbf{p}},{\mathbf{v}}_{\mathrm{P}})&=\sum_{r=1,t=1}^{r=R,t=T}\rho_{r,t}^2\frac{|(\boldsymbol{\psi}^{\mathrm{(xy)}}_{r,t})^{\mathsf{H}}\mathbf{y}_{r,t}|^2}{E^2_{r,t}},
	\end{aligned}	
\end{equation}
which is consistent with the weights naturally occurring in the proposed fusion framework. As a result, we have demonstrated that the SNR-aware weights in Bayesian probability fusion framework achieve maximum global SNR.

\vspace{-0.3cm}
\subsection{Convergence of PCGA Algorithm}
\vspace{-0.3cm}

We employ the fixed-point theorem for proving the convergence of the PCGA algorithm. According to \eqref{eq:CGA1}, for the iterative process of the PCGA algorithm, we define the mapping $G(\mathbf{p})$ as
\begin{equation}\label{eq:cproof}
	\begin{aligned}	
		G(\mathbf{p})=\mathbf{p}+\eta\nabla{ f_1(\mathbf{p})},
	\end{aligned}
\end{equation}
where $\nabla{ f_1(\mathbf{p})}$ is the gradient of the fused spacial pseudo-spectrum $f_1({\mathbf{p}})$ at $\mathbf{p}$. Let $\mathbf{p}^*$ denote the solutions that satisfies $G(\mathbf{p}^*)=\mathbf{p}^*$.

First, we prove the existence of the fixed point for $G(\mathbf{p})$. Specifically, $f_1({\mathbf{p}})$ can be written as
\begin{equation}\label{eq:cproof1}
	\begin{aligned}		f_1({\mathbf{p}})=\sum_{r=1,t=1}^{r=R,t=T}\sum_{l=1}^{l=L}\frac{\rho_{r,t}^2g_{r,t,l}(\mathbf{p})}{(K+1/\rho_{r,t}^2)},
	\end{aligned}
\end{equation}
where $g_{r,t,l}(\mathbf{p})=|(\boldsymbol{\psi}^{\mathrm{(xyf)}}_{r,t})^{\mathsf{H}}\mathbf{y}_{r,t}^{\mathrm{f}}(l)|^2$ can be regarded as a Fourier transform process, transforming signal from the frequency domain to the range domain. Recall ${y}_{r,t}(k,l)$ is expressed as
\begin{equation}\label{eq:cproof3}
	\begin{aligned}	
		{y}_{r,t}(k,l)=\left\{
		\begin{aligned}
			\tilde{\beta}_{r,t}\mathrm{e}^{-{\mathrm j}2\uppi k\Delta_{\mathrm{f}} \frac{d_{r,t}({\mathbf{p}}^*)}{c}}+z(l),\quad &0\leq k \leq K-1; \\
			0,\qquad\qquad\qquad &\text{otherwise}.\\
		\end{aligned}
		\right.
\end{aligned}
\end{equation}
According to the Fourier transform pair, $g_{r,t,l}(\mathbf{p})$ is written as
\begin{equation}\label{eq:cproof4}
\begin{aligned}	
	g_{r,t,l}(\mathbf{p})=\left|\sum_{k=-\infty}^{+\infty}\mathrm{e}^{{\mathrm j}2\uppi k \Delta_{\mathrm{f}} \frac{d_{r,t}(\mathbf{p})}{c}}{y}_{r,t}(k,l)\right|^2=\left|\frac{\sin(2\uppi\Delta_{\mathrm{f}} K(d_{r,t}(\mathbf{p})-d_{r,t}({\mathbf{p}}^*))/2c)}{\sin(2\uppi\Delta_{\mathrm{f}} (d_{r,t}(\mathbf{p})-d_{r,t}({\mathbf{p}}^*))/2c)}\right|^2+\varepsilon.
\end{aligned}
\end{equation}
Let $\Delta d_{r,t}(\mathbf{p})=d_{r,t}(\mathbf{p})-d_{r,t}({\mathbf{p}}^*)$, which can be proved as a convex function. Moreover, $g_{r,t,l}(\Delta d_{r,t}(\mathbf{p}))$ is similar to the sinc function and can be proved as an approximate convex function within the main lobe $-\frac{c}{2K\Delta_{\mathrm{f}}}\leq\Delta d_{r,t}(\mathbf{p})\leq\frac{c}{2K\Delta_{\mathrm{f}}}$, where $\frac{c}{2K\Delta_{\mathrm{f}}}$ is the resolution of range estimation.
By utilizing the properties of convex functions, if $g_{r,t,l}(\mathbf{p})$ is convex for all t/rAP pairs, it can be proved that $f_1({\mathbf{p}})$ is convex as well. Thus, the fused spacial pseudo-spectrum $f_1({\mathbf{p}})$ is a local convex function and has a single extreme point for ${\mathbf{p}}\in\mathbb{P}_0$. This indicates that there exists a fixed point ${\mathbf{p}^*}\in\mathbb{P}_0$ satisfying $G({\mathbf{p}}^*)={\mathbf{p}}^*$.

Next, we demonstrate that the PCGA algorithm converges to this fixed point. According to Banach fixed point theorem, taking the limit of $\mathbf{p}^{(i+1)}$ yields
\begin{equation}\label{eq:cproof7}
\begin{aligned}	\lim_{i\rightarrow\infty}\mathbf{p}^{(i+1)}=\lim_{i\rightarrow\infty}G(\mathbf{p}^{(i)})\Longleftrightarrow\lim_{i\rightarrow\infty}\mathbf{p}^{(i)}={\mathbf{p}}^*.
\end{aligned}
\end{equation}
This implies that the update process of the PCGA algorithm converges the fixed point ${\mathbf{p}}^*$.

The proposed PCGA algorithm restricts the initial value of the iteration within the neighborhood of the maximum by means of a coarse search, ensuring that the value of position is within the main lobe of the function $g_{r,t,l}(\mathbf{p})$. As a result, the convergence of the algorithm is guaranteed.

\vspace{-0.3cm}
\subsection{CRLB of Parameter Estimation}
\label{APPB}
\vspace{-0.3cm}

The CRLB of position estimation can be derived from the CRLB of bistatic range estimation. Let $\mathbf{d}\in\mathbb{R}^{N\times 1}$ be the vector consisting of the bistatic ranges of all t/rAP pairs, where $N=RT$ and ${d}_n=d_{r,t}$, $r=n-R\left\lfloor(n-1)/R\right\rfloor$, $t=1+\left\lfloor(n-1)/R\right\rfloor$. Based on (\ref{eq:backprojection}a), the Jacobian matrix $\mathbf{J}^{\mathrm{(d)}}={\partial \mathbf{d}}/{\partial \mathbf{p}}\in\mathbb{R}^{N\times 2}$ of the bistatic range with respect to the target position for each T/R AP pair is computed as
\begin{align}\label{eq:aa}
	J_{n,1}^{\mathrm{(d)}}=\frac{\partial d_n}{\partial x}=\frac{x-x_r}{||\mathbf{p}-\mathbf{p}_r||_2}+\frac{x-x_t}{||\mathbf{p}-\mathbf{p}_t||_2}\quad\textrm{and}\quad
	J_{n,2}^{\mathrm{(d)}}=\frac{\partial d_n}{\partial y}=\frac{y-y_r}{||\mathbf{p}-\mathbf{p}_r||_2}+\frac{y-y_t}{||\mathbf{p}-\mathbf{p}_t||_2}.
\end{align}
According to \cite{2022AdhamNear}, the CRLB of bistatic range estimation is $\delta d^2_{n}=\delta d^2_{r,t}=\frac{3\sigma^2}{\uppi^2 (\Delta_{\mathrm{f}}/c)^2\tilde{\sigma}_{r,t}^2K(K^2-1)L}$. With the CRLB of range estimation and the Jacobian matrix, the Fisher information matrix is written as
\begin{align}\label{eq:aa1}
\mathbf{I}(\mathbf{p})=(\mathbf{J}^{\mathrm{(d)}})^{\mathsf{T}}(\mathbf{\Pi}^{\mathrm{(d)}})^{-1} \mathbf{J}^{\mathrm{(d)}}=\sum_{n=1}^{n=N}\frac{1}{\delta d^2_{n}}\left[\begin{matrix}
	(J_{n,1}^{\mathrm{(d)}})^2 &J_{n,1}^{\mathrm{(d)}}J_{n,2}^{\mathrm{(d)}}\\
	J_{n,2}^{\mathrm{(d)}}(J_{n,1}^{\mathrm{(d)}})&(J_{n,2}^{\mathrm{(d)}})^2
\end{matrix} \right],
\end{align}
where $\mathbf{\Pi}^{\mathrm{(d)}}=\mathrm{diag}\left\{[\delta d^2_{1},\cdots,\delta d^2_{n},\cdots,\delta d^2_{N}]\right\}$. Thus, the CRLB of position estimation is $\mathrm{CRLB}(\mathbf{p})=\mathbf{I}^{-1}(\mathbf{p})$.

The CRLB of velocity estimation can be derived in a similar manner. Let $\mathbf{v}\in\mathbb{R}^{N\times 1}$ be the vector consisting of the speeds of all t/rAP pairs. Based on (\ref{eq:backprojection}b), the Jacobian matrix $\mathbf{J}^{\mathrm{(v)}}={\partial \mathbf{v}}/{\partial \mathbf{v}_{\mathrm{P}}}\in\mathbb{R}^{N\times 2}$ of the speed with respect to the target velocity for each T/R AP pair is derived as
\begin{align}\label{eq:aa}
	J_{n,1}^{\mathrm{(v)}}=\frac{\partial v_n}{\partial v_x}=\frac{x-x_r}{||\mathbf{p}-\mathbf{p}_r||_2}+\frac{x-x_t}{||\mathbf{p}-\mathbf{p}_t||_2}\quad\textrm{and}\quad
	J_{n,2}^{\mathrm{(v)}}=\frac{\partial v_n}{\partial v_y}=\frac{y-y_r}{||\mathbf{p}-\mathbf{p}_r||_2}+\frac{y-y_t}{||\mathbf{p}-\mathbf{p}_t||_2}.
\end{align}
According to \cite{2022AdhamNear}, the CRLB of speed estimation is $\delta v^2_{n}=\delta v^2_{r,t}=\frac{3\sigma^2}{\uppi^2 (T_{\mathrm{P}}/\lambda)^2\tilde{\sigma}_{r,t}^2L(L^2-1)K}$. With the CRLB of speed estimation and the Jacobian matrix, the Fisher information matrix is also written as $\mathbf{I}(\mathbf{v}_{\mathrm{P}})=(\mathbf{J}^{\mathrm{(v)}})^{\mathsf{T}}(\mathbf{\Pi}^{\mathrm{(v)}})^{-1} \mathbf{J}^{\mathrm{(v)}}$, where $\mathbf{\Pi}^{\mathrm{(v)}}=\mathrm{diag}\left\{[\delta v^2_{1},\cdots,\delta v^2_{n},\cdots,\delta v^2_{N}]\right\}$. Thus, the CRLB of velocity estimation is $\mathrm{CRLB}(\mathbf{v}_{\mathrm{P}})=\mathbf{I}^{-1}(\mathbf{v}_{\mathrm{P}})$.

\section{Numerical Simulation and Discussions}
\label{sec:simulations}
\vspace{-0.3cm}

In this section, we provide numerical simulation results to show the performance of our proposed fusion framework and PCGA algorithm. To better show the superiority of our proposal, we adopt the following schemes for comparison:
\newline(a) {{\textbf{Signal fusion}}}: The scheme involves summing the matched-filtered signals from multiple t/rAP pairs, specifically employing coherent/non-coherent BP algorithms \cite{2022AdhamNear};
\newline(b) {{\textbf{Parameter fusion (hard)}}}: The scheme involves formulating a system of equations with the estimated distances and velocities from multiple t/rAP pairs, specifically employing the weighted least squares algorithm \cite{2022ZhangEfficient};
\newline(c) {{\textbf{Parameter fusion (soft)}}}: The scheme involves first estimating the range and velocity, then reconstructing the probability function, and finally conducting the fusion, which is a suboptimal form of Bayesian probability fusion \cite{2024FigueroaCooperative};
\newline(d) {\textbf{{Symbol fusion}}}: The scheme involves multiplying the real parts of phase-compensated signals from multiple t/rAP pairs \cite{2024WeiSymbol}.

\begin{figure}[h]
	\vspace{-0.4cm}
	\centering
	\begin{minipage}{0.43\linewidth}
		\centering
\includegraphics[width=1.03\linewidth]{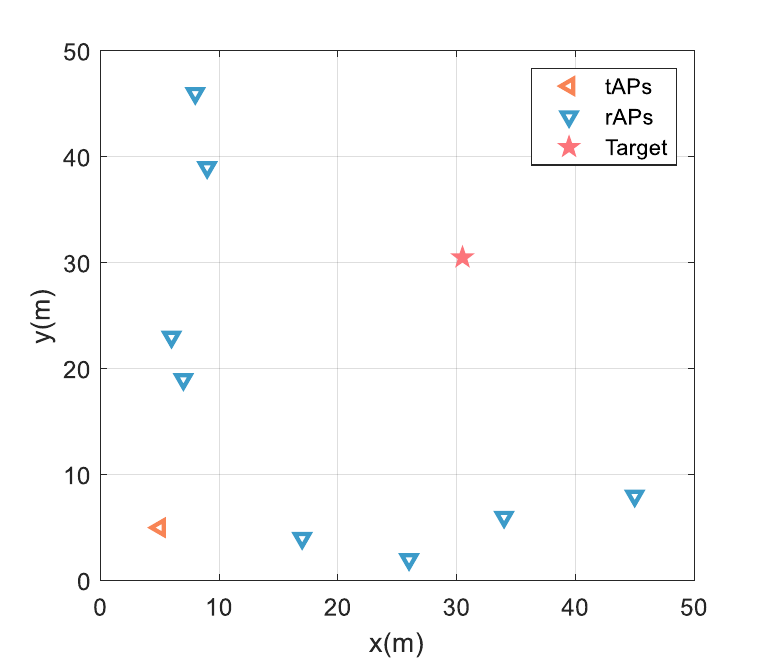}
\vspace{-0.6cm}
\caption{Position of rAPs, tAPs and target in the test.}
\label{fig:topo}
	\end{minipage}
	%\qquad
	\begin{minipage}{0.56\linewidth}
		\centering
		\includegraphics[width=1.02\linewidth]{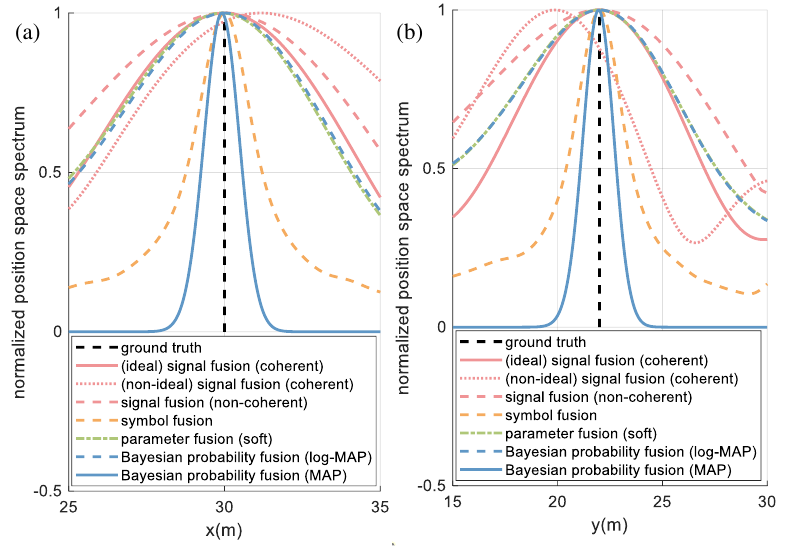}
		\vspace{-0.8cm}
		\caption{Cross-section of generated $2$D spatial pseudo-spectrum. (a) coordinate $x$. (b) coordinate $y$.\protect\footnotemark[3]}
		\label{fig:xy_spectrum}
	\end{minipage}
		\vspace{-0.4em}
\end{figure}

We consider one tAP ($T=1$) and eight rAPs ($R=8$), and all APs are connected to a CPU via fronthaul links in $2$D space. As shown in Fig. \ref{fig:topo}, the positions of tAP are ($5.0$, $5.0$) meters (m), and the positions of the rAPs are ($17$, $4$) m, ($7.0$, $19.0$) m, ($45.0$, $8.0$) m, ($26.0$, $2.0$) m, ($9.0$, $39.0$) m, ($6.0$, $23.0$) m, ($8.0$, $46.0$) m, and ($34.0$, $6.0$) m, respectively. The positions of the target position are set as ($30.5$, $30.5$) m with velocity ($2.7$, $2.0$) meters per second (m/s). In the simulation, the tAP transmits the sensing signals, and after being reflected by the target, the sensing signals are received by the rAPs. We consider a typical millimeter wave system with analog beamforming structure, the carrier frequency and the subcarrier spacing are set as $f_c=30$ GHz and $\Delta_{\mathrm{f}}=240$ KHz, respectively; the symbol repetition interval is set as $T_p=0.625\;{\rm{ms}}$; the number of subcarriers is set as $K=100$ and the number of symbols is set as $L=100$. Each AP employs an array with $M=64$ elements and the antenna spacing is half-wavelength.

\vspace{-0.1cm}
\footnotetext[3]{Since it is difficult to ensure the coherence of signals in real-world systems, the following signal fusion only refers to the signal fusion (non-coherent). For single target scenario, the sharpness of spectrum does not affect the accuracy of parameter estimation, thus Bayesian probability fusion (MAP) and Bayesian probability fusion (log-MAP) are considered equivalent hereinafter.}

\vspace{-0.3cm}
\subsection{Results of Spatial Pseudo-Spectrum}
\vspace{-0.3cm}

First, we compared the cross-section of $2$D spatial pseudo-spectrum obtained by various fusion methods. As shown in Fig. \ref{fig:xy_spectrum}, (ideal) signal fusion (coherent), signal fusion (non-coherent), symbol fusion, parameter fusion (soft), Bayesian probability fusion (log-MAP) and Bayesian probability fusion (MAP) can all generate a unimodal spectrum of $2$D position.
From the perspective of the accuracy of the spectrum, an analysis is carried out as following.
It can be observed that signal fusion (non-coherent), Bayesian probability fusion (log-MAP), Bayesian probability fusion (MAP), parameter fusion (soft) and symbol fusion can all obtain the spectrum peak near the ground truth. Specifically, when the signals of each t/rAP pair satisfy ideal coherence, signal fusion (coherent) can obtain the accurate spectrum peak. However, it is difficult for each r/rAP pair to meet the coherence condition. In the case of non-ideal signals, signal fusion (coherent) will cause the spectrum peak to deviate from the true value. In addition, an analysis is carried out from the perspective of the sharpness of the spectrum. Compared with other algorithms, the spectrum generated by the proposed Bayesian probability fusion (MAP) is the sharpest. The sharpness of the spectrum generated by symbol fusion ranks second, yet it is superior to that of signal fusion.

\begin{figure*}[t]
	\vspace{-0.4cm}
\centering
\includegraphics[width=1.0\linewidth]{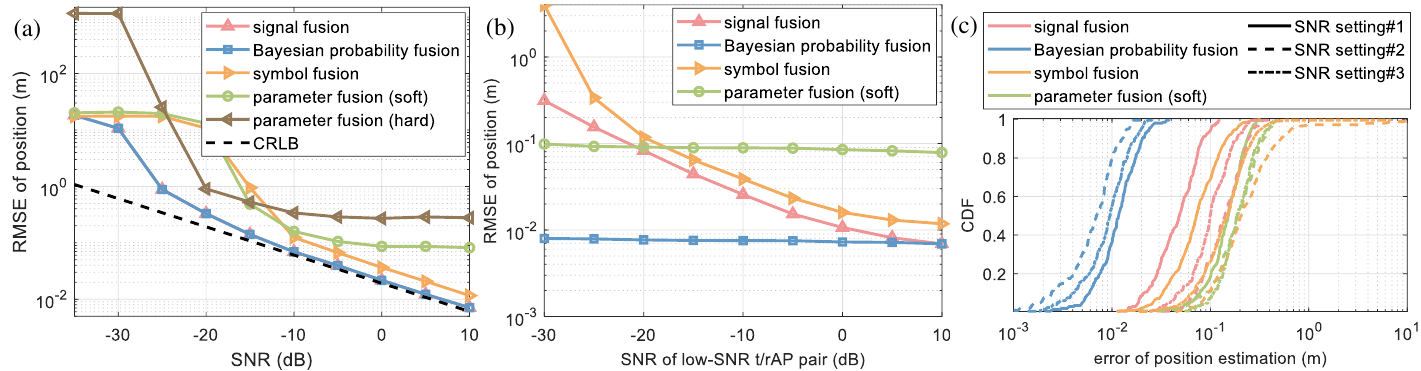}
	\vspace{-0.6cm}
\caption{Comparison of position estimation performance. (a) RMSE vesus SNR. (b) RMSE vesus SNR of low-SNR t/rAP pair. (c) CDF of position estimation error for different SNR settings.\protect\footnotemark[4]}
\label{fig:pos_rmse1}
\vspace{-0.3em}
\end{figure*}

\begin{figure*}[t]
\centering
\includegraphics[width=1.0\linewidth]{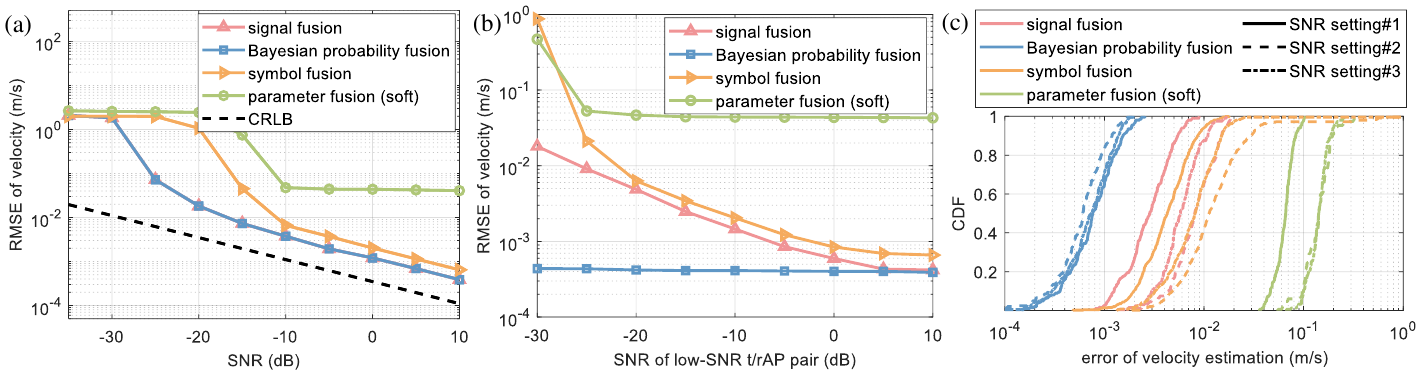}
	\vspace{-0.6cm}
\caption{Comparison of velocity estimation performance. (a) RMSE vesus SNR. (b) RMSE vesus SNR of low-SNR t/rAP pair. (c) CDF of velocity estimation error for different SNR settings.}
\label{fig:v_pos_rmse1}
\vspace{-0.2cm}
\end{figure*}

\footnotetext[4]{Given that the performance of parameter fusion (hard) is far from being on par with other methods in Fig. \ref{fig:pos_rmse1}, it will no longer be adopted as a comparative method in the subsequent analysis.}

\vspace{-0.3cm}
\subsection{Parameters Estimation Error}
\vspace{-0.3cm}

To further compare the accuracy of different fusion methods in parameters estimation, the root mean square error (RMSE) of position and velocity are defined as
\begin{align}\label{eq:RMSE}	
	\mathrm{RMSE}_{\mathrm{P}}=\sqrt{\mathbb{E}\left\{ ( x_0-\hat{x})^2+( y_0-\hat{y})^2 \right\}}\quad\textrm{and}\quad
	\mathrm{RMSE}_{\mathrm{V}}=\sqrt{\mathbb{E}\left\{ ( v_{x,0}-\hat{v}_x)^2+( v_{y,0}-\hat{v}_y)^2 \right\}},
\end{align}
where $(x_0,y_0)$, $(\hat{x},\hat{y})$, $(v_{x,0},v_{y,0})$ and $(\hat{v}_x,\hat{v}_y)$ are the ground truth of position, the estimated position, the ground truth of velocity and the estimated velocity, respectively. We compared RMSE of position and velocity estimation by signal fusion, the proposed Bayesian probability fusion, symbol fusion, parameter fusion (soft) and parameter fusion (hard). Meanwhile, we derive the CRLB \cite{2022AdhamNear} of position and velocity estimation as a benchmark in Section \ref{APPB}.

\begin{figure}[t]
	\vspace{-0.4cm}
	\centering
	\begin{minipage}{0.53\linewidth}
		\centering
		\includegraphics[width=1\linewidth]{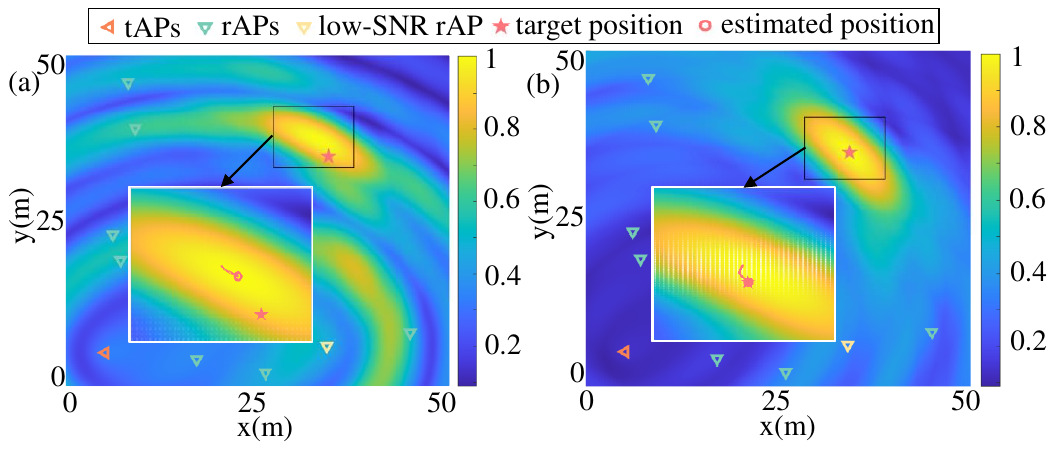}
		\vspace{-0.6cm}
		\caption{PCGA search processes of $2$D spatial pseudo-spectrum with one low SNR (-10dB) t/rAP pair and seven high SNR (10dB) pair. (a) Signal fusion. (b) Bayesian probability fusion.}
		\label{fig:mapxy}
		\vspace{-0.8cm}
	\end{minipage}
	\begin{minipage}{0.46\linewidth}
		\centering
		\includegraphics[width=1\linewidth]{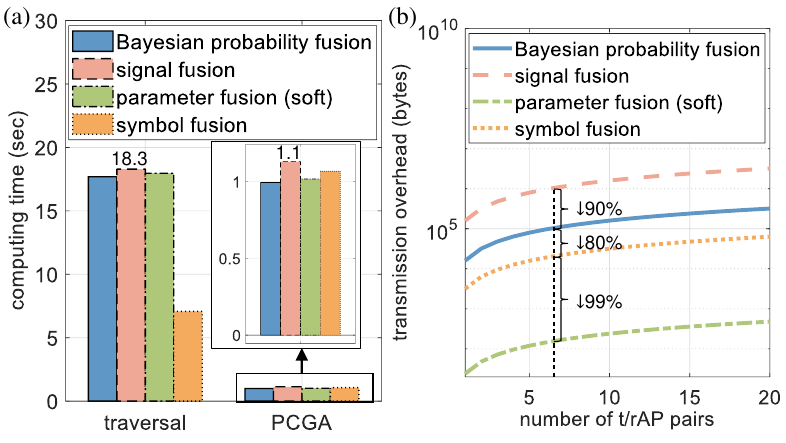}
		\vspace{-0.6cm}
		\caption{Computational complexity and overhead of different fusion modes. (a) Runtimes of different fusion modes solved by traversal and PCGA algorithms. (b) Transmission overhead of different fusion modes.}
		\label{fig:time}
		\vspace{-0.2cm}
	\end{minipage}
\end{figure}

As shown in Fig.~\ref{fig:pos_rmse1}(a), the RMSE of position estimated by each fusion method decreases as the SNR increases. Moreover, the performance of the proposed fusion is similar to signal fusion, and both are superior to symbol fusion, parameter fusion (soft) and parameter fusion (hard). In low SNR region, the performance of parameter fusion (hard) is the worst, the performances of parameter fusion (soft) and symbol fusion are close. This is because all the fusion algorithms become ineffective in low SNR region.
In high SNR region, the performance of parameter fusion (soft) is inferior to the signal fusion, Bayesian probability fusion and symbol fusion. This is because parameter fusion (soft) requires estimation of the ranges for each t/rAP pair, which will introduce range estimation errors for position estimation. Similarly, the compensation of symbol fusion requires velocity estimation, which also introduces errors.

In order to evaluate the advantages of the Bayesian probability fusion, we compared the performance of different fusion algorithms under the condition that the SNR of each t/rAP pair is varying. To control the comparison conditions, we fixed the number of low SNR t/rAP pairs to one and changed their SNR from $-30\;{\rm{dB}}$ to $10\;{\rm{dB}}$. The comparison results are shown in Fig. \ref{fig:pos_rmse1}(b). The performance of signal fusion and symbol fusion improves as the SNR of the low SNR t/rAP pair increases, while the proposed fusion and parameter fusion (soft) fluctuates slightly. Signal fusion sums the moduli of the matched filtering results of each t/rAP pair. As a result, it is affected by the low SNR t/rAP pairs to a greater extent than our proposed fusion and parameter fusion (soft) with SNR-aware weighted square sum. In addition, we present a cumulative distribution function (CDF) comparison of different fusion methods under three SNR settings, including SNR setting$\#1:$ $(10,10,10,10,-10,-10,-10,-10)\;{\rm{dB}}$, SNR setting$\#2:$$(20$,$20$, $10$,$10$,$-10$,$-10$,$-20$,$-20)\;{\rm{dB}}$, and SNR setting$\#3:$$(20,15,10,5,-5,-10,-15,-20)\;{\rm{dB}}$. These SNR settings reflect the general situations in real-world scenarios. As shown in Fig.~\ref{fig:pos_rmse1}(c), the performance of our proposed fusion is optimal under each SNR setting. As shown in Figs.~\ref{fig:v_pos_rmse1}(a)-(c), the performance of each method in velocity estimation is consistent with that in position estimation, showing similar trends.

To clearly demonstrate the advantages of Bayesian probability fusion, we present the PCGA search processes of signal fusion and our proposed fusion in Fig.~\ref{fig:mapxy}(a) and Fig.~\ref{fig:mapxy}(b), respectively. It can be observed that when there exist a low SNR ($-10\;{\rm{dB}}$) t/rAP pair among other high SNR ($10\;{\rm{dB}}$) pairs, the spectrum peak of signal fusion is shifted, and the PCGA algorithm fails to obtain the true target position. In contrast, Bayesian probability fusion can converge to the true target position.

\vspace{-0.3cm}
\subsection{Computational Complexity and Transmission Overhead}
\vspace{-0.3cm}

Finally, we compare the computational complexity and the transmission overhead of different fusion modes. Regarding computing efficiency, we statistically analyzed the time required for different fusion modes solved by traversal and PCGA algorithm through $1000$ Monte Carlo trials on Intel(R) Core(TM) i7-13700K CPU. In the trials, we set the SNR as $10\;{\rm{dB}}$, $L = 100$, $K = 100$, and number of t/rAP pairs as $8$. As shown in Fig. \ref{fig:time}(a), the computing time of each fusion model solved by traversal will be higher than that by PCGA algorithm. Besides, due to the preprocessing of subcarrier dimension and symbol dimension being completed, the computing time required for symbol fusion in traversal search is lower. Nevertheless, after adopting PCGA algorithm, the computing time of various fusion modes has been reduced to the same magnitude. For example, the computing time of signal fusion is reduced from $18.3\;{\rm{s}}$ to $1.1\;{\rm{s}}$ after adopting PCGA algorithm. Furthermore, based on the distributed architecture, we compared the transmission overhead of each fusion model. We set the SNR as $10\;{\rm{dB}}$, $L = 100$, $K = 100$, and statistically analyzed the variation of transmission overhead with the number of t/rAP pairs. As shown in Fig.~\ref{fig:time}(b), the transmission overheads are in descending order of: signal fusion, Bayesian probability fusion, symbol fusion and parameter fusion (soft). Compared with signal fusion, the transmission overhead of our proposed framework is reduced by $90\%$. This is because signal fusion require the transmission of channel state information to the fusion center. In contrast, Bayesian probability fusion, symbol fusion and parameter fusion (soft) can utilize the processing capabilities of distributed APs to reduce the amount of data. In detail, Bayesian probability fusion only needs to transmit signals within specific ranges and velocities by utilizing priori information, symbol fusion only needs to transmit the real parts of phase-compensated signals, and soft parameter fusion only needs to transmit the estimated ranges and velocities.

\section{Real-World Experiments}
\label{sec:experiment}
\vspace{-0.3cm}

To provide additional validation of our proposed fusion framework and evaluate its parameters estimation performance in a complex and realistic environment, we conducted experiments of unmanned aerial vehicle (UAV) intrusion detection in an outdoor testing field.

\begin{figure*}[h]
	\vspace{-0.2em}
	\centering
	\includegraphics[width=0.75\linewidth]{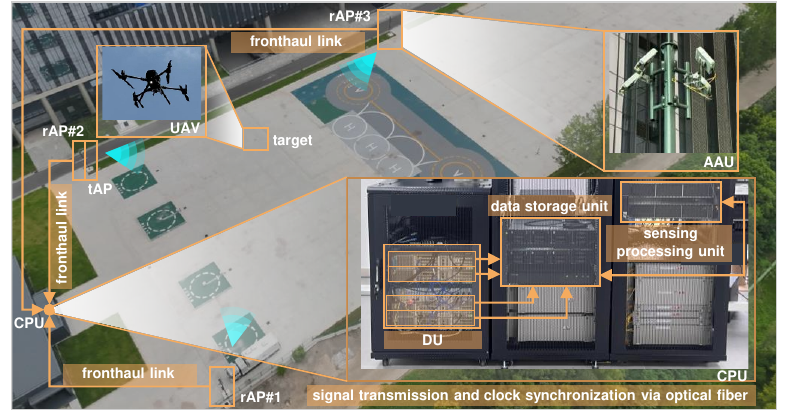}
	\vspace{-0.5em}
	\caption{Experimental environment and hardware setups for field test.}
	\label{fig:scenarios}
	\vspace{-0.5cm}
\end{figure*}

Fig.~\ref{fig:scenarios} provides a photograph of the experimental environment and depicts the setup of the measurement system. The outdoor test field, with its complex distribution of scatterers (e.g., high-rise buildings, dense vegetation), effectively reproduces the realistic multipath propagation characteristics of urban areas. To set up the experimental configuration for data collection, we installed commercial mmWave active antenna units (AAUs) on pole-mounted brackets in the test field. These AAUs are connected to the distributed units (DUs) via optical fibers to enable signal transmission and clock synchronization. After completing the baseband processing, the DUs transmit the raw signals to the data storage unit, and the sensing processing unit reads the signals from the data storage unit to perform the sensing function. Under this experimental configuration, the AAUs act as APs, the DUs, data storage unit and sensing processing unit serve as CPU. As a target, the UAV is equipped with centimeter-level real-time kinematic (RTK) positioning function, which provides reliable ground truth data for conducting meticulous comparisons and performance evaluations. The settings of experimental parameters are listed in Table. \ref{table:setting}. In the experiment, we configured four AAUs as one tAP and three rAPs respectively, which were placed in fixed positions ($0.0$, $0.0$, $13.7$) m, ($0.0$, $67.2$, $13.1$) m, ($0.0$, $0.0$, $13.7$) m, ($90.9$, $0.3$, $13.3$) m. While the UAV flied from ($-96.7$, $23.0$, $114.0$) m to ($-38.9$, $23.0$, $114.0$) m with velocity ($8.0$, $0.0$, $0.0$) m/s, all APs performed beam scanning. As such, $7.3$ seconds of raw signals were collected for performance evaluation.\protect\footnotemark[5]

\begin{table}[h]
	\centering
	\caption{Experiment settings.}
	\label{table:setting}
	\small
	\begin{tabular}{c|c||c|c}
		\hline
		& & &\\[-15pt]
		parameter&value&parameter&value\\
		\hline
		& & &\\[-15pt]
		central frequency $f_{\mathrm{c}}$&$25.6\;\mathrm{GHz}$&number of subcarriers $K$&$1024$\\
		& & &\\[-15pt]
		subcarrier spacing $\Delta_{\mathrm{f}}$&$120\;\text{KHz}$&transmit power $P_{\mathrm{T}}$&$45\;\mathrm{dBm}$\\
		& & &\\[-15pt]
		bandwidth $B$&$200\;\text{MHz}$&number of symbols $L$&$512$\\
		& & &\\[-15pt]
		symbols repetition interval $T_{\mathrm{P}}$&$0.156\;$ms&RCS of UAV $\gamma$& $\sim0.01\;\text{m}^2$\\
		\hline
	\end{tabular}
\end{table}

\begin{figure*}[]
	\vspace{-0.2em}
	\centering
	\includegraphics[width=0.99\linewidth]{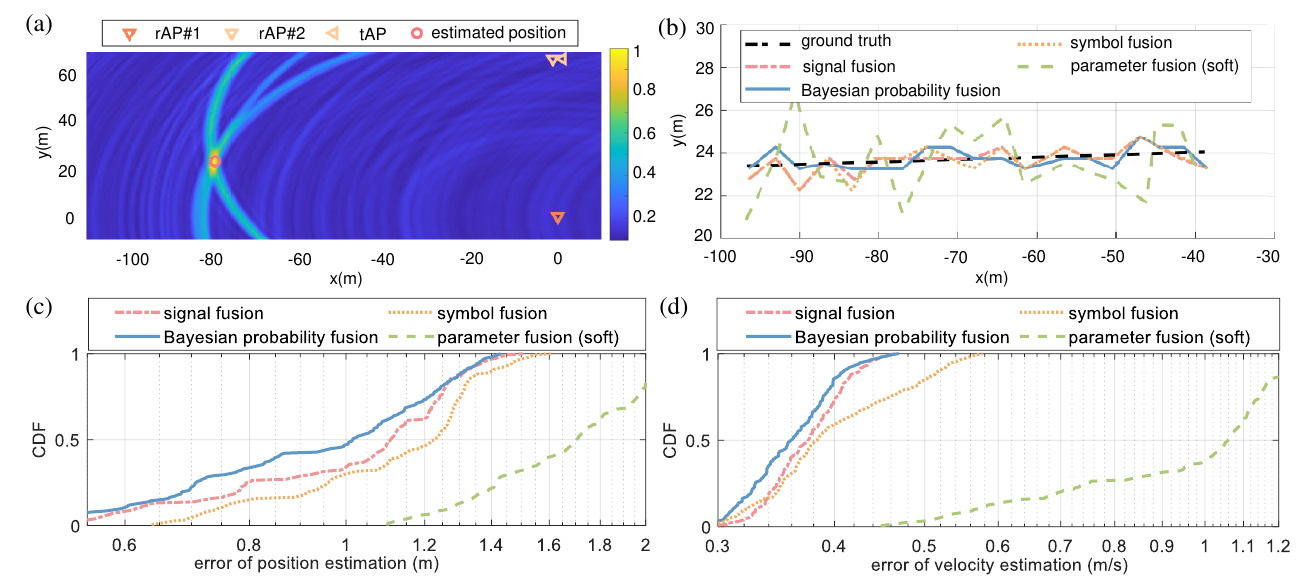}
	\vspace{-0.6em}
	\caption{Experiment results in an outdoor testing field. (a) $2$D spatial pseudo-spectrum generated by Bayesian probability fusion. (b) Estimated target trace. (c) CDF of position estimation error. (d) CDF of velocity estimation error.}
	\label{fig:UAV_map}
\end{figure*}

After performing operations such as clutter filtering and beam selection, we completed the performance evaluation of parameter estimation. As shown in Fig.~\ref{fig:time}(a), under the condition of fixed altitude, the proposed fusion can generate the expected 2D spatial pseudo-spectrum, which can be used to search for the maximum value via PCGA. Fig.~\ref{fig:time}(b) shows the target traces estimated by various methods as well as the ground truth (true value), from which it can be seen that the estimated traces are roughly the same as the ground truth in terms of overall trend. The CDF curves of position and velocity estimation are illustrated in Fig.~\ref{fig:UAV_map}(c) and Fig.~\ref{fig:UAV_map}(d). Considering 2D position estimation with a fixed altitude, approximately $50\%$ error of parameter fusion (soft), symbol fusion, signal fusion and  the proposed method are less than $1.7$ m, $1.2$ m, $1.1$ m, $1.0$ m, respectively. Thus, our proposed method can reduce the error by $41\%$ compared to parameter fusion (soft), and reduce the error by $9\%$ compared to signal fusion with lower transmission overhead. Due to the existence of AP position deviations and clock offsets in real-world scenarios, the experimental results are inferior to the simulation results. Although Bayesian probability fusion only leads the other methods by a slight advantage, it can play a more significant role in scenarios where the SNR difference among APs is larger and the number of APs is more sufficient.

\section{Conclusion}
\label{sec:Conclusion}
\vspace{-0.3cm}

In this paper, we proposed a Bayesian probability fusion framework to enhance the performance of networked collaborative sensing. On the one hand, within the Bayesian paradigm, the parameter estimation problem is transformed into a constrained MAP optimization problem.
On the other hand, the PCGA algorithm was developed to address the computational challenges of the high-dimensional problem.
Furthermore, theoretical contributions including the derivation of optimal weighting scheme for global SNR maximization, rigorous convergence analysis of the PCGA algorithm, and the establishment of the CRLB for the proposed estimator, provides fundamental insights and apply to diverse fusion schemes.
Numerical simulation and real-world experiment results validated the superiority of the proposed framework, which shows that it achieves $90\%$ reduction in transmission overhead relative to signal fusion and exhibits $41\%$ lower estimation error compared to parameter fusion. These findings highlight the potential of Bayesian probability fusion strategies in practical multistatic sensing systems, particularly for scenarios with varying channel states among multi-AP networks. Future work will explore parameter estimation in multi-target scenarios, as well as calibration schemes for the presence of systematic biases.

\Acknowledgements{This work was supported in part by the National Science and Technology Major Project under Grant No. 2024ZD1300200 and the Fundamental Research Funds for the Central	Universities under Grant No. 2242023R40005.}


\begin{thebibliography}{99}

	\bibitem{2022ITUR}
	ITU-R. Future technology trends of terrestrial international mobile telecommunications systems towards 2030 and beyond. Report M.2516-0. 2022.

	\bibitem{2023YouTowards}
	You X H, Huang Y M, Liu S H, et al. Toward 6G TK$\mathrm{\mu}$ extreme connectivity: Architecture, key technologies and experiments. {IEEE Trans Wireless Commun}, 2023, 30: 86-95.
	
	\bibitem{2022IMT}
	IMT-2030. 6G Typical Scenarios and Key Capabilities White Paper. IMT-2030 (6G) Promotion Group, 2022.
	
	\bibitem{3GPP}
	 3GPP. Study on scenarios and requirements for next generation access technologies. TR.38.913. https://www.3gpp.org/ftp/Specs/archive/38\_series/38.913/38913-i00.zip

	\bibitem{2021YouTowards}
	You X H, Wang C X, Huang J, et al. Towards 6G wireless communication networks: vision, enabling technologies, and new paradigm shifts. {Sci China Inf Sci}, 2021, 64: 1-74.

	
\bibitem{2024Gonz}
	Gonz{\'a}lez-Prelcic N, and Keskin M F, Kaltiokallio O, et al. The integrated sensing and communication revolution for 6G: Vision, techniques, and applications.  {Proceedings of the IEEE}, 2024, 112: 676-723.
	
	\bibitem{2022LiuIntegrated}
	 Liu F, Cui Y H, Masouros C, et al. Integrated sensing and communications: Toward dual-functional wireless networks for 6G and beyond. {IEEE J Sel Areas Commun}, 2022, 40: 1728-1767.

	\bibitem{TOIT25}
	Liu S H, Fu N N, Zhang Z H, et al. Integrated user scheduling and beam steering in over-the-air federated learning for mobile IoT. {ACM Trans Internet Techn}, 2025, in press.
	
	\bibitem{2025LiuFeasibility}
	Liu G Y, Xi R Y, Jiang T, et al. Feasibility study of cooperative sensing: radar cross section, synchronization, cooperative cluster, performance and prototype. {Sci China Inf Sci}, 2025, 68: 1-19.

	\bibitem{2024LiuISAC}
	Liu S H, Gao S T, Hong Y X, et al. ISAC-oriented access point placement in cell-free mMIMO systems. Electron Lett, 2024, 60: e70037.

	\bibitem{2024XuAccess}
	Xu F F, Liu S H, Mao Z H, et al. Access point deployment for localizing accuracy and user rate in cell-free systems. In: Proceedings of ACM Annual International Conference on Mobile Computing And Networking (MobiCom), 2024. 2148-2154.
	
	\bibitem{2023WangFull}
	Wang D M, You X H, Huang Y M, et al. Full-spectrum cell-free RAN for 6G systems: System design and experimental results. {Sci China Inf Sci}, 2023, 66: 1-14.

	\bibitem{2023CaoExperimental}
	Zeng F, Liu R Y, Sun X Y, et al. Multi-static ISAC based on network-assisted full-duplex cell-free networks: performance analysis and duplex mode optimization. {Sci China Inf Sci}, 2025, 68: 1-22.
	\bibitem{2024YangCoordinated}
	Yang X Y, Wei Z Q, Xu J, et al. Coordinated transmit beamforming for networked ISAC with imperfect CSI and time synchronization. {IEEE Trans Wireless Commun}, 2024, 23: 18019-18035.
	\bibitem{2025ZhangSignal}
	Zhang Z H, Zhu J H, Liu N, et al. Signal fusion method for networked radar motivated by data fusion. {IEEE Sens J}, 2025: 951-961.
	
	\bibitem{2022AdhamNear}
	Sakhnini A, Bast S D, Guenach M, et al. Near-field coherent radar sensing using a massive MIMO communication testbed. {IEEE Trans Wireless Commun}, 2022, 21: 6256-6270.
	
\bibitem{2025Favarelli}
	Favarelli E, Matricardi E, Pucci L, et al. Sensor fusion and resource management in MIMO-OFDM joint sensing and communication. {IEEE Trans Veh Techn}, 2025, 74: 9284-9298.

	\bibitem{2025LiuModel}
	Liu S H, Mao Z H, Li X K, et al. Model-driven deep neural network for enhanced direction finding with commodity 5G gNodeB. {ACM Trans Sens Netw}, 2025, 21: 13.
	
	\bibitem{2014Yin}
	Yin P L, Yang X P, Liu Q H, et al. Wideband distributed coherent aperture radar. In: {Proceeding of IEEE Radar Conference (RadarConf)}, 2014. 1114-1117.
	
	\bibitem{2023XiongDistributed}
	Xiong K, Cui G L, Yi W, et al. Distributed localization of target for MIMO radar with widely separated directional transmitters and omnidirectional receivers. {IEEE Trans Aerosp Electron Syst}, 2023, 59: 3171-3187.
	
	\bibitem{2024VermaTrack}
	Verma J K, Chhabra J K, Ranga V. Track consensus-based labeled multi-target tracking in mobile distributed sensor network. {IEEE Trans Mob Comput}, 2024, 23: 7351-7362.
	

	\bibitem{2024MoussaMulti}
	Moussa A, Liu W, Zhang Y D, et al. Multi-target location and doppler estimation in multistatic automotive radar applications. {IEEE Trans Radar Sys}, 2024, 2: 215-225.
	\bibitem{2022ZhangEfficient}
	Zhang X D, Wang F Z, Li H B. An efficient method for cooperative multi-target localization in automotive radar. {IEEE Signal Process Lett}, 2022, 29: 16-20.

	\bibitem{2024WeiSymbol}
	Wei Z Q, Xu R Z, Feng Z Y, et al. Symbol-level integrated sensing and communication enabled multiple base stations cooperative sensing. {IEEE Trans Veh Technol}, 2024, 73: 724-738.
	
	
	\bibitem{2025LiuHRT}
	Liu S H, Yan H, Li X K, et al. HRT-Net: A cell-free cooperative sensing method based on attention-weighted hierarchical network (in Chinese). Journal of Radars, 2025, 14: 974:993.
	
	\bibitem{2021RenImproved}
	Ren M L, He P, Zhou J J. Improved shape-based distance method for correlation analysis of multi-radar data fusion in self-driving vehicle. {IEEE Sens J}, 2021, 21: 24771-24781.

	\bibitem{2021Yi}
    Yi W and Chai L. Heterogeneous multi-sensor fusion with random finite set multi-object densities. {IEEE Trans Signal Process}, 2021, 69: 3399-3414.

	\bibitem{2022KolianderFusion}
	Koliander G, El-Laham Y, Djuri\'{c} P M, et al. Fusion of probability density functions. {Proc IEEE}, 2022, 110: 404-453.
	

	\bibitem{2021GaoReliable}
	Gao B, Jia M, Zhang T T, et al. Reliable target positioning in complicated environments using multiple radar observations. In: {Proceeding of IEEE Global Communication Conference (GLOBECOM)}, 2021. 1-6.

	\bibitem{2025Potter}
	Potter M, Akcakaya M, Necsoiu M, et al. Multistatic-radar RCS-signature recognition of aerial vehicles: A Bayesian fusion approach. {IEEE Trans Aerospace Electron Syst}, 2025, 61: 151-161.

	\bibitem{2024FigueroaCooperative}
	Figueroa M R, Bishoyi P K, Petrova M. Cooperative multi-monostatic sensing for object localization in 6G networks. In: {Proceeding of IEEE Wireless Communication Networking Conference (WCNC)}, 2024. 1-6.
	
	\bibitem{2016HeathOverview}
	Heath R W, Gonz\'{a}lez-Prelcic N, Rangan S, et al. An overview of signal processing techniques for millimeter wave MIMO systems.  {IEEE J Sel Top Signal Process}, 2016, 10: 436-453.
	\bibitem{2017SuhConstruction}
	Suh J, Kim C, Sung W, et al. Construction of a generalized DFT codebook using channel-adaptive parameters. {IEEE Commun Lett}, 2017, 21: 196-199.
	
	\bibitem{2023Wei5G}
	Wei Z Q, Wang Y, Ma L, et al. 5G PRS-based sensing: A sensing reference signal approach for joint sensing and communication system. {IEEE Trans Veh Technol}, 2023, 72: 3250-3263.
	
	\bibitem{2022Chen2022}
	Chen W R, Li L X, Chen Z, et al. Enhancing THz/mmWave network beam alignment with integrated sensing and communication. {IEEE Commun Lett}, 2022, 26: 1698-1702.
	
	\bibitem{1960SwerlingProbability}
	Swerling P. Probability of detection for fluctuating targets. {IRE Trans Inf Theory}, 1960, 6: 269-308.
	
\bibitem{2024HuaMIMO}
	Hua H C, Han T X, Xu J. MIMO integrated sensing and communication: CRB-rate tradeoff. {IEEE Trans Wireless Commun}, 2024, 23: 2839-2854.
	
\bibitem{2024ZhaoBayesian}
	Zhao L Q, Yan L, Duan X J, et al. A Bayesian multistage fusion model for radar antijamming performance evaluation. {IEEE Trans Aerosp Electron Syst}, 2024, 60: 729-740.
	
\bibitem{1950Sherman}
	Sherman J, Morrison W J. Adjustment of an inverse matrix corresponding to a change in one element of a given matrix. {Ann Math Statist}, 1950, 21: 124-127.
	
\bibitem{2020ZhouSpatial}
	Zhou Y, Xu D Z, Tu W L, et al. Spatial information and angular resolution of sensor array. {Signal Process}, 2020, 174: 1-10.
	
\bibitem{2022SakhniniTarget}
	Sakhnini A, Guenach M, Bourdoux A, et al. A target detection analysis in cell-free massive MIMO joint communication and radar systems. In: {Proceeding of IEEE International Conference on Communication (ICC)}, 2022. 2567-2572.

	\bibitem{2025YangGradient}
	Yang Y Q, Wang W M, Zhang F C, et al. Gradient-based beam peak search for over-the-air testing of mmWave phased arrays. {IEEE Antennas Wireless Propagat Lett}, 2025, 24: 157-161.
	
\end{thebibliography}
\end{document}